\documentclass{rspublic21}
\usepackage{bbm}
\usepackage{amsfonts}
\usepackage{amssymb}
\usepackage[numbers,sort&compress]{natbib}
\usepackage{graphicx}
\usepackage{amsmath}
\usepackage{appendix}
\usepackage{amsthm}
\usepackage{epigraph}
\usepackage{xcolor}
\usepackage[a4paper,top=2in,bottom=1in,left=0.7in,right=1.2in]{geometry}

\input{tcilatex}
\begin{document}

 \renewcommand{\emph} [1] {{#1}}

\title{On the effect of decoherence on quantum tunnelling. }
\author[A.Y. Klimenko]{A.Y. Klimenko \thanks{%
email: klimenko@mech.uq.edu.au}}
\affiliation{The University of Queensland, SoMME, QLD 4072, Australia \\ \bigskip \small{\it (SN Applied Sciences, 2021, 3:710) }}

\maketitle

\begin{abstract}
{decoherence, quantum tunnelling, non-equilibrium dynamics} \textbf{Abstract}
This work proposes a series of quantum experiments that can, at least in
principle, allow for examining microscopic mechanisms associated with
decoherence. \emph{ These experiments can be interpreted as a quantum-mechanical
version of non-equilibrium mixing between two volumes separated by a thin
interface. } One of the principal goals of such experiments is in identifying
non-equilibrium conditions when time-symmetric laws give way to
time-directional, irreversible processes, which are represented by
decoherence at the quantum level. The rate of decoherence is suggested to be
examined indirectly, with minimal intrusions --- this can be achieved by
measuring tunnelling rates that, in turn, are affected by decoherence.
Decoherence is understood here as a general process that does not involve
any significant exchanges of energy and governed by a particular class of
the Kraus operators. The present work analyses different regimes of
tunnelling in the presence of decoherence and obtains formulae that link the
corresponding rates of tunnelling and decoherence under different
conditions. \emph{ It is shown that the effects on tunnelling of intrinsic decoherence and of decoherence 
due to unitary interactions with the environment are similar 
but not the same and can be  distinguished in experiments. } \ 
\end{abstract}

\section{Introduction \label{Sec1}}

The goal of this work is to consider experiments that can, at least in
principle, examine time-directional quantum effects in an effectively
isolated system. Such experiments need to be conducted somewhere at the
notional boundary between the microscopic quantum\ and macroscopic
thermodynamic worlds, that is we need to deal with quantum systems that can
exhibit some degree of thermodynamic behaviour. At the quantum level, this
corresponds to persisting decoherence, which is, perhaps, the most
fundamental irreversible process that we are aware of --- it takes place at
the smallest scales, increases entropy \cite{Abe2020} and, expectedly,
induces various macroscopic effects associated with the thermodynamic arrow
of time \cite{mixing2020}. A large volume of literature is dedicated to
decoherence, which may involve both intrinsic \cite{Zurek2002,
QTreview,Beretta2005,Stamp2012} and environmental \cite%
{Zurek1982,Joos1984,Joos2003,EnvDec2005,CT-G2006,CT-P2006,Stamp2012,Yukalov2011}
mechanisms.

{\ The present work examines a problem that, at least conceptually, can
become an experiment probing the direction of time.\ This problem represents
a quantum-mechanical version of non-equilibrium mixing between two volumes
separated by a thin interface. In this quantum version of the classical
problem, particles tunnel through the interface and, at the same time, are
subject to the omnipresent influence of quantum decoherence, which,
presumably, is the fundamental mechanism enacting non-equilibrium,
time-directional effects in the macroscopic world \cite{SciRep2016}. }

In quantum experiments, one has to face another fundamental difficulty ---
interferences from the environment and measurements. Environmental
interferences can overwhelm intrinsic mechanisms of decoherence, while\
quantum measurements routinely cause decoherences and collapses (which are
interpreted here as defined in the Appendix of Ref. \cite{Klimenko2019})
instead of observing these decoherences and collapses without interfering.
It appears, however, that, under the conditions examined in this work,
decoherence affects the rate of quantum tunnelling and, therefore, can be
characterised by the tunnelling rates without measuring decoherence
directly. Among many formulations of tunnelling problems \cite%
{LL3,Tunn2003,Dattagupta2004,Tunn2009,TunnZeno2014,Tunn2017,QTG2018}, we
select one that has a transparent and, at the same time, sufficiently
general solution. For this formulation involving quantum tunnelling through
a high potential barrier under non-equilibrium conditions, we examine
mechanisms that may be ultimately responsible for the direction of time.
Conducting such experiments is not easy but seems possible even under the
current level of technology. {\ Conceptually and technically similar
experiments have been performed in the past \cite%
{ImryYoseph2002,Dattagupta2004,Dec1Exp2008,TunnZeno2014,QTG2018}. These
experiments investigated mesoscopic decoherence in context of the
Aharonov-Bohm effect \cite{ImryYoseph2002}, proton tunnelling under thermal
bath conditions \cite{Dattagupta2004}, the effect of invasive frequent
measurements on quantum tunnelling \cite{TunnZeno2014} (i.e. the \textit{%
quantum Zeno effect} \cite{Zeno1977}).} Our main interest is in examining
decoherence by using tunnelling but, unlike in the previous of experiments,
under conditions that avoid direct interferences from the environment and
measurements, and screen the experiment from a supposed direct influence of
the temporal boundary conditions imposed on the universe.
\footnote{The key question investigated in the present work --- whether the intrinsic and environmental mechanisms of decoherence can be distinguished in experiments --- has been repeatedly raised in the past: for example, by G.N. Hatsopoulos and G.P. Beretta in "Where is the entropy challenge?"  [AIP Conf. Proc. 1033, 34-54 (2008)].}

This manuscript is organised as follows. Section \ref{Sec2} briefly reviews
the interpretation of the arrow of time from a philosophical perspective
pointing to the ideas of Hans Reichenbach as a source of critical thinking
about the time that is relevant to the present work. The readers, who are
interested in quantum mechanics and non-equilibrium dynamics more than in
philosophical issues, can omit this section at first reading. Section \ref%
{Sec3} introduces the tunnelling problem and, in the context of this
problem, discusses emergence of the arrow of time. Section \ref{Sec4}
reviews different time-asymmetric and time-symmetric interpretations of
quantum mechanics, in particular the two-state vector formalism \cite%
{2S-QM1964,2state1998,2S-QM2008}. Section \ref{Sec5} examines tunnelling in
absence of decoherence, while Section \ref{Sec6} investigates the influence
of decoherence on the tunnelling rates. Section \ref{Sec7} discusses conduct
of experiments based on the results of this work. Section \ref{Sec8}
summarises our conclusions. More extensive derivations of asymptotic
tunnelling rates are presented in \ref{SecA} and a brief consideration of
the problem from the perspective of the theory of environmental decoherence 
\cite{Zurek1982} is given in \ref{SecB}.

\section{Discrimination of the past and the future from a philosophical
perspective \label{Sec2}}

It is well known that the most important physical laws --- those of
classical, relativistic and quantum mechanics --- are time-symmetric, but
our experience of physical reality strongly discriminates the past and the
future. The observed arrow of time is reflected in the second law of
thermodynamics, which permits entropy increases but bans reduction of
entropy in isolated thermodynamic systems. While the Boltzmann time
hypothesis, which suggests that the perceived arrow of time and the
direction of entropy increase must be the same (i.e. connected at some
fundamental level), may be striking at first, but after some thinking over
the issue, most people tend to arrive\ at the same conclusion. Since Ludwig
Boltzmann \cite{Boltzmann-book,Klimenko2019}, the overall conditions
prevailing in the universe (or its observable part) have been thought to be
responsible for this temporal asymmetry. In modern physics, the increasing
trend for entropy is commonly explained by the asymmetry of temporal
boundary conditions imposed on the universe, i.e. by low-entropy conditions
at the time of Big Bang \cite{PenroseBook}. This explanation is called the 
\textit{past hypothesis }by Albert \cite{Albert2000} and in other
publications. There are no doubts that the past conditions existing in the
universe are very important. The pertinent question, however, is not whether
these conditions are important, but whether the direct influence of the
initial conditions imposed on the universe is sufficient to explain all
time-directional phenomena observed in experiments. A number of publications
seem to be content with the sufficiency of the special initial conditions in
the early universe to explain all entropy increases in thermodynamically
isolated systems, even if it is presumed that all laws of physics are
time-symmetric \cite{Albert2000,North2011}.

The alternative view is that the past hypothesis is important but, on its
own, is insufficient to fully explain entropy increases required by the
second law. This view can be traced to the \textit{principle of parallelism
of entropy increase}, which was introduced by Hans Reichenbach \cite%
{Reichenbach1971}, and further explained, evaluated and extended by Davies 
\cite{Davies1977}, Sklar \cite{Sklar1993} and Winsberg \cite%
{Winsberg2004a,Winsberg2004b}. The Reichenbach principle concurs that
initial conditions imposed on the universe can explain many effects
associated with entropy increase; nor does it deny that entropy can
fluctuate. {\ The initial conditions imposed on the universe may explain why
entropy tends to increase more often than to decrease in semi-isolated
thermodynamic subsystems (branches) but, assuming that all governing
physical laws are time-symmetric, these initial conditions do not explain
the persistence and consistency of this increase (this, of course, does not
exclude the existence of fluctuations of entropy but indicates that,
according to the fluctuation theorem, entropy increases over a given time
interval are consistently more likely than entropy decreases \cite%
{Searles2008}).} Consider a system that is isolated from the rest of the
universe without reaching internal equilibrium: would such a system
demonstrate conventional thermodynamic behaviour, or would its entropy
increase terminate under these conditions? Reichenbach \cite{Reichenbach1971}
conjectured that such an isolated system would still display conventional
thermodynamic properties, and we do not have any experimental evidence to
the contrary. The principle of parallelism of entropy increase is useful not
only as a thought experiment. When applied at a physical level,
Reichenbach's ideas lead us to the existence of a time priming mechanism
that continues to exert its influence even in isolated conditions \cite%
{KM-Entropy2014,Klimenko2019,mixing2020}. Implications of the directionality
of time in quantum mechanics \cite{Abe2020} and chemical kinetics \cite%
{Maas2020} are further discussed in special issue \cite{Entropy2021}.

Huw Price \cite{PriceBook} pointed out that our temporal (antecedent)
intuition often results in implicit discrimination of the directions of time
in physical theories --- this tends to introduce conceptual biases that may
be difficult to identify due to the all-encompassing strength of our
intuitive perception of time. These biases conventionally involve
assumptions associated with the conceptualisation of antecedent causality,
such as imposing\ initial (and not final) conditions or presuming stochastic
independence before (and not after) interactions. These assumptions are very
reasonable and supported by our real-world experience, but may form a
logical circle: effectively, we often presume antecedent causality in order
to explain entropy increase forward in time, which, in turn, is used to
explain and justify antecedent causality \cite{mixing2020}. Here, we
recognise that the laws of classical and quantum mechanics are
time-symmetric, and will endeavour to avoid implicit discrimination of the
directions of time \cite{Klimenko2019}. When reading this article, the
reader, who is accustomed to thinking in terms of antecedent causality,
might feel that something is strange or missing. The concept of time priming
used here is aimed at avoiding intuitive assumptions introducing
directionality of time by implying antecedent causality in one form or
another --- time priming does not seem relevant whenever antecedent
causality is presumed. Since the directions of time are, obviously, not
equivalent, there must be a physical mechanism that is responsible for this
and, at least in principle, testable in experiments. One of such possible
mechanisms, pointing to interactions of quantum effects and gravity, has
been suggested by Penrose \cite{Penrose1996}. Another possibility is that
this mechanism is related to the temporal asymmetry of matter and antimatter 
\cite{K-PhysA, KM-Entropy2014,SciRep2016,Ent2017}. In the present work,
however, we do not presume any specific form of the mechanism and use this
special term, the \textit{time primer,} as a place holder for possible
physical explanations. Detecting the time primer in experiments is most
likely to be difficult due to the expected smallness of its magnitude.

The Reichenbach parallelism principle is not a trivial statement or
tautology: one can imagine a state of affairs in which this principle has
only limited validity. A thermodynamic system, placed in isolation and
screened by equilibrium states from the initial and final conditions imposed
on the universe, might, at least in principle, cease to exhibit
thermodynamic, entropy-increasing behaviour even if non-equilibrium
conditions are created within a selected time interval. \textit{%
Reichenbach's conjecture} tells us that this should not happen: such an
isolated system would still tend to increase its entropy similarly to and in
parallel with entropy increases of various thermodynamic systems scattered
in the rest of the universe. While general implications of the Reichenbach
principles are discussed in Ref. \cite{mixing2020}, our broad goal is to
consider specific experiments where these principles can be examined
directly or indirectly but, desirably, examined in a way that can give some
indications of the underlying mechanisms responsible for decoherence and,
ideally, for the direction of time. While the experiments suggested in the
present work are related to modern quantum mechanics more than to
Reichenbach's branch model, one needs to acknowledge that these experiments
are following the direction of his thinking.

\section{Quantum mixing in a branch system \label{Sec3}}

{\ This section introduces a detailed description of the problem, which, as
noted above, represents a quantum-mechanical version of mixing across an
interface that is deemed to be branched and isolated from the rest of the
universe. }

\subsection{Formulation of the problem}

\begin{figure}[h]
\begin{center}
\includegraphics[width=12cm,page=1,trim=3cm .3cm 1.5cm 2cm, clip ]{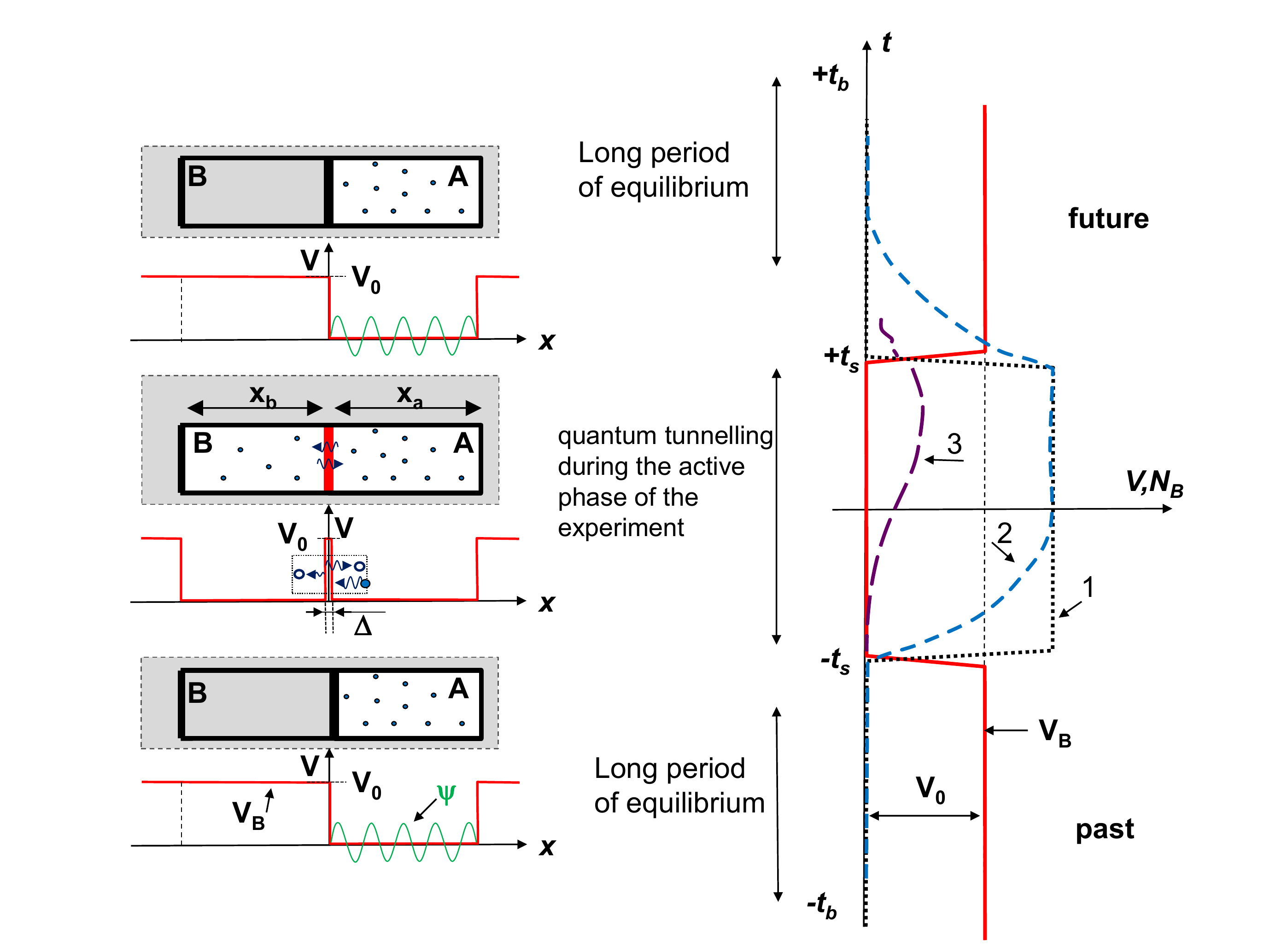}
\caption{System, which is isolated from the environment and screened from the
temporal boundary conditions imposed on the universe by equilibrium states,
involves quantum tunnelling from section A to section B and back in response
to the time-symmetric disturbance of the potential $V_{\text{B}}$. Here, $N_{%
\text{B}}$ is expected number of particles in section B: 1 - under
equilibrium, 2 - as predicted by master equations with dominant decoherence
and by thermodynamic considerations;  3 - possible quantum solution without decoherence.   
}
\label{fig1}
\end{center}
\end{figure}

We consider a number (say, $N_{0}$) of quantum particles placed in a
rectangular box AB, which is partitioned into two sections A and B. The
quantum levels in the system are sparsely populated so that the rules of the
Maxwell-Boltzmann statistics apply. The tunnelling particles are initially
located in section A of the box AB, as shown in Figure \ref{fig1}. The value
of the potential $V$ is prohibitively high in section B to permit any
significant presence of the particles in this section. Section A also
contains a number $N_{1}$ of inert, non-tunnelling particles, and this
number is sufficiently large so that the system of particles can be expected
to behave thermodynamically ($N_{1}\gg N_{0}$). We expect that all of the
particles achieve thermodynamic equilibrium during the passive stages of the
experiment (the particles interact with each other but these interactions
are deemed to be weak). The particles in the box are trapped by a potential
field and completely isolated from the environment (which is also deemed to
be in an equilibrium state) during the duration of the experiment $%
-t_{b}<t<+t_{b}.$ In a more simple experimental setup, the tunnelling
particles\ can be brought into thermodynamic equilibrium with their
container box. Using inert particles, however, allows us to control the
statistical scale of the experiment.

The equilibrium state is maintained for a long time $\sim t_{b}$\ prior to
(and after) the active phase of the experiment --- this time is much larger
than the characteristic thermalisation time $\tau _{t}$ for this system $%
t_{b}\gg \tau _{t}$. Thermalisation implies achieving thermodynamic
equilibrium between all particles under consideration while, in the present
context, equilibration implies reaching steady-state concentrations of the
tunnelling particles. Depending on conditions, equilibration may or may not
require thermalisation. During the passive stages of the experiment,
equilibration requires energy exchanges between different modes and,
therefore, presumes thermalisation. In the active stage of the experiment,
however, the system may reach equilibrium values of $N_{\text{A}}$ and $N_{%
\text{B}}$ without reaching (or without significantly disturbing)
thermodynamic equilibrium. Thermalisation and equilibration are generally
not synonymous \cite{Yukalov2011}. Thermalisation requires a substantial
energy exchange between different modes to reach the equilibrium
thermodynamic distributions, while this is not necessarily the case for
equilibration.

The active phase of the experiment $-t_{s}\leq t\leq +t_{s}$ is short $%
t_{s}\ll t_{b}$. At time $t=-t_{s},$ the particle access to section B of the
box is opened by rapid lowering of the potential $V(\mathbf{r},t)$ in this
section to the same level as in section A, so that the tunnelling particles
can now tunnel through the barrier that separates sections A and B, while
the inert particles remain in section A. The rate of change of the potential
is fast compared to the characteristic tunnelling time so that the
concentrations of the particles in sections A and B deviate from their
equilibrium values. Particles tunnel from A to B and back through a
potential barrier separating the sections until the process is terminated at 
$t=t_{s}$ in a time-symmetric manner by increasing potential in section B to
its original value. The experiment is expected to follow by a long-lasting
equilibrium state, where particles are again in a thermodynamic equilibrium
(this implies that their location is in section A). It is also worthwile to
consider the temporal symmetry in the experiment of lowering and rising the
potential $V(\mathbf{r},t)$ in a piston-like manner (e.g. as discussed in
Refs. \cite{mixing2020,AYK2020}), although the present work is primarily
focused on examining the effects of decoherence on tunnelling.

\subsection{Expected emergence of temporal asymmetry}

Would the concentration of particles in section B behave in a thermodynamics
manner with time-delayed relaxations towards its equilibrium value as
illustrated by curve 2 in Figure \ref{fig1}? The Reichenbach conjecture
states that it would: despite being completely isolated from the environment
and fully screened by the equilibrium states from the initial and final
conditions imposed on the universe, the system is still expected to display
time-directional thermodynamic behaviour (this, of course, needs to be
confirmed by actual experiments, but, for the sake of the argument, we
assume at this point that the Reichenbach conjecture is correct). If the
laws of the universe are time-symmetric, the initial and final conditions
are similar, and interactions of the system and the universe take place only
through strictly time-symmetric disturbance of the potential $V(\mathbf{r}%
,t)=V(\mathbf{r},-t)$, why the response of the system to time-symmetric
inputs is evidently time-asymmetric?

\subsection{Environmental interferences}

While we have declared complete isolation of the system, controlling and
eliminating environmental interferences for a particle or a system of
particles may, in real-world experiments, be quite difficult. According to
thinking common among many physicists, any quantum system is always subject
to influences of the environment; environmental interactions, no doubts, can
cause and do cause decoherences as indicated in many theories and
experiments \cite%
{Zurek1982,Joos2003,EnvDec2005,CT-G2006,CT-P2006,Stamp2012,Yukalov2011,Nature2020decoherence}%
. The interferences that involve a measurable, specific influence of
environment on the system (such as the direct effect of cosmic radiation on
superconducting qubits measured in Ref. \cite{Nature2020decoherence}) can be
evaluated and, at least in principle, protected from in these experiments.
Bell entanglement of two elementary particles can be protected from
environmental interferences and preserved for a very long time. However, the
presumed omnipresent environmental interference that induces decoherence but
does not have any specific measurable mechanism and any specific physical
particles or surrounding objects casing it (e.g. interference involving a
special quantum field that is like, say, the Higgs field present
everywhere), is, conceptually, no different from intrinsic decoherence.
While the distinction between the intrinsic and environmental mechanisms of
decoherence is blurred and depends on exact definitions, the principal
difference between various theories of decoherence and thermalisation is in
relying or not relying on time-symmetric scientific frameworks (such as
unitary evolutions in quantum mechanics). We therefore distinguish intrinsic
or effectively intrinsic mechanisms of decoherence from unitary
interferences with the environment.

Unitary environmental interferences do not, by themselves, discriminate the
directions of time: all theories of environmental decoherence based on
time-symmetric physical laws (e.g. unitary evolutions of quantum mechanics)
must involve another principal element --- an assumption that violates the
equivalence of the directions of time. The effect of the environment on
decoherence becomes clear only if we presume antecedent causality (although
causality is something that we have vowed to avoid in the present work).
Indeed, in the absence of directionality of time required by antecedent
causality, environmental interactions can induce recoherences in the same
way as they induce decoherences. The time-directional effect of random
interventions produced by the environment is determined by imposing initial
(as opposite to final) conditions on the system (individual realisations of
a random process with independent increments are not time-directional \cite%
{mixing2020}). Note that the immediate environment can be in a thermal
equilibrium state and experience only time-symmetric fluctuations or, even
better, be kept at (nearly) absolute zero temperature (while\ radiation that
can transmit interactions over long distances is expected to be
decoherence-neutral by itself \cite{Ent2017}). In the casual model, the
final state depends on random interventions but the initial state does not,
because the initial state is fixed but the final state is not. The
interventions are deemed to cause the final state but not the initial state.
If we fix the final state instead of fixing the initial state, then the
effect of environmental interferences would be recohering. By themselves,
the environmental interferences only introduce some effective randomness
into the system. Presuming antecedent causality is the central element of
the major theories of environmental interference --- it is causality and not
the interference that breaks the symmetry of the directions of time.

\subsection{Initial and final conditions imposed on the universe}

One, of course, may abandon antecedent causality and, instead, invoke
low-entropy initial conditions imposed on the universe. While these
conditions must be very important, the main question concerning our
experiment remains: how can these conditions influence the stochastic state
of the system after a long period of equilibrium? We may assume that $%
N_{0}\sim 1$ so that the system of $N_{1}$ inert particles under
consideration is small (while $t_{b}$ is extremely long, much much longer
than $t_{s}$) and repeatedly experiences very substantial fluctuations
around equilibrium during the passive phases of the experiment. Due to these
fluctuations, we may, in principle, select the time moments $t=\pm t_{b}$
when the final state at $t=+t_{b}$ has lower entropy than the initial state
at $t=-t_{b}$ --- does this mean that the arrow of time should be reversed
during the active phase of the experiment and the changes in $N_{\text{B}}$
-- the number of particles in section B -- would tend to preempt the changes
of the potential $V(\mathbf{r},t)$ rather than to follow them? While the
negative answer to this question is expected, the physical mechanism that
can allow the temporal boundary conditions imposed on the universe to affect
the active phase of our experiment is not obvious. Or, alternatively, should
the arrow of time disappear and the directions of time become equivalent
under these conditions? According to the Reichenbach conjecture, we tend to
believe that the arrow of time should persist.\ 

In the suggested experiment, the system cannot preserve any statistical
information about the conditions that preceded the experiment or follow the
experiment --- equilibrium states achieve maximal entropy and necessarily
destroy all such information. Yet, there must be a physical mechanism that
discriminates the directions of time in lieu of the direct action of the
temporal boundary conditions imposed on the universe if the Reichenbach
conjecture is correct. This mechanism is called here the \textit{time primer}%
. Conceptually, the time primer does not replace global temporal boundary
conditions imposed on the universe but reflects the local action of these
global conditions.

The time primer may act predominantly on a larger system and propagate to a
semi-localised subsystem through time-priming interference, which (the
subsystem) in this case can behave similarly to the effect of intrinsic time
priming but without any intrinsic time priming on its own. One may invoke
time priming in larger and larger environments but this interpretation
neither gives a complete explanation (now we need to explain time priming in
the environment, which may well be in its equilibrium state) nor helps the
experiments (instead of confining and measuring the effect of interest, we
disperse it over the environment in a way that it is likely to become
experimentally untraceable). Therefore, environmental interactions should be
avoided as much as possible or, at least, they need to be measured and
quantified. The boundary conditions imposed on the universe may indeed
determine the direction of decoherence in a tiny experiment with quantum
mixing, but this influence must have a specific physical mechanism and
should be measurable and quantifiable.

\subsection{Conditions of the experiment}

By changing the parameters of the experiment, we can observe different
physical conditions. The system of particles may consist of one or more
particles, which do not strongly interact between\ themselves and may or may
not interact with a thermodynamic (or statistical microscopic) object (e.g.
a system of inert particles), while the particles and the object remain
fully isolated from the environment. The characteristic tunnelling time can
also be changed by varying the shape of the potential. If the active phase
of the experiment is sufficiently short $t_{s}\ll \tau _{t}$, then there is
no substantial exchange of energy takes place within the system during the
active phase. This, however, does not imply that quantum particles evolve
unitarily since they may still be affected by decoherence. Since the
equilibrated system should be, from the quantum perspective, in or close to
its maximally mixed state under specified conditions (or in the effectively
maximally mixed state specified by canonical typicality \cite%
{CT-G2006,CT-P2006}), thermalisation necessarily implies decoherence, and,
therefore, the characteristic decoherence time cannot be longer than the
characteristic thermalisation time. In fact, one may expect the decoherence
time $\tau _{d}$ to be significantly shorter than the thermalisation time $%
\tau _{t}$ (at least for sufficiently large systems) so that characteristic
time $\tau _{d}$ can be shorter or comparable to $t_{s},$ even if $t_{s}$ is
much smaller than $\tau _{t}.$ In the case of\ $\tau _{d}\ll t_{s},$
decoherence must have a strong influence on the experiment. If however, the
active phase of the experiment is much shorter than the characteristic
decoherence time $t_{s}\ll \tau _{d}$, then decoherence has little effect on
the quantum system of particles, which is now expected to evolve unitarily
and be governed by the Schr\"{o}dinger equation during the active phase.

\section{Time-directional and time-symmetric interpretations of quantum
mechanics\label{Sec4}}

This section discusses time-asymmetric and time-symmetric interpretations of
quantum mechanics. The former tends to imply antecedent causality, while the
latter can be used to avoid implicit discrimination of the directions of
time.

\subsection{Schr\"{o}dinger equation and its solution}

As we have to deal with quantum mixtures, different particles generally do
not form coherent superpositions, and quantum wave functions and density
matrices are more useful tools than quantum fields under these conditions.
Hence, we focus first our attention on behaviour of a single quantum
particle but remember that interactions between particles within the system
are conducive to decoherence. According to the conventional interpretation
of the problem, evolution of the wave function is governed by the \textit{%
Schr\"{o}dinger equation} in the position representation 
\begin{equation}
i\hbar \frac{\partial \psi }{\partial t}=\mathbb{H}\psi ,\ \ \ \ \ \mathbb{H}%
=\frac{1}{2m}\mathbbm{p}^{2}+\mathbb{V}=-\frac{\hbar ^{2}}{2m}\nabla ^{2}+V(%
\mathbf{r},t)  \label{1Shr}
\end{equation}%
with the Dirichlet (zero) boundary conditions 
\begin{equation}
\psi =0\ \text{at }\mathbf{r}\in \partial \text{AB,\ \ \ AB}=\text{A}\cup 
\text{B}  \label{1bc}
\end{equation}%
since the potential $V$ is assumed to be very high at and beyond the
boundaries. Here, $\psi $ is the wave function, $t$ is time, $\mathbb{H}$ is
the Hamiltonian, $\hbar $ is the Planck constant and $m$ is the particle
mass. Relations (\ref{1Shr}) and (\ref{1bc}) apply to all wave functions
that correspond to different particles, assuming that interactions between
particles can be neglected. The sections A and B are separated by a thin
high-energy barrier located near $x=0.$\ The probability of tunnelling is
relatively small but essential; tunnelling may or may not be affected by
decoherence as discussed in the rest of this paper.

According to the formulation of the problem presented above, the potential $V
$ is assumed to be time-independent $V(\mathbf{r},t)=V(\mathbf{r})$ within
the time interval $-t_{s}<t<+t_{s}$, which is of interest in the present
work. Since the Hamiltonian is Hermitian $\left\langle \phi \middle|\mathbb{H%
}\psi \right\rangle =\left\langle \mathbb{H}\phi \middle|\psi \right\rangle $%
, the solution of the problem is based on the Hilbert--Schmidt theorem 
\begin{equation}
\psi (t^{\circ },\mathbf{r})=\mathbb{U(}t^{\circ })\psi _{0}=\exp \left( 
\frac{\mathbb{H}}{i\hbar }t^{\circ }\right) \psi _{0}=\sum_{j}\psi
_{j}=\sum_{j}a_{j}\exp \left( -i\omega _{j}t^{\circ }\right) \Psi _{j}(%
\mathbf{r})  \label{1sol}
\end{equation}%
where 
\begin{equation}
a_{j}=\frac{\left\langle \Psi _{j}\middle|\psi _{0}\right\rangle }{Q_{j}},\
\ \ \mathbb{H}\Psi _{j}=E_{j}\Psi _{j},\ \ \ \left\langle \Psi _{j}\middle%
|\Psi _{i}\right\rangle =\delta _{ji}Q_{j},\ \ \ \omega _{j}=\frac{E_{j}}{%
\hbar }  \label{eig1r}
\end{equation}%
$\psi _{0}=\left. \psi \right\vert _{t=t_{0}}$ specifies the initial
conditions, $t^{\circ }=t-t_{0},$\ and the energy eigenstates $\Psi _{j}(%
\mathbf{r})$ satisfy the same boundary conditions as $\psi $. The initial
(or final) conditions can be set at $t_{0}=-t_{s}$\ or at $t_{0}=+t_{s}$.
The jumps of the potential at $t=\pm t_{s}$ are assumed to be so rapid that
the wave function does not have time to adjust and $\left. \psi \right\vert
_{t=t_{0}+0}=\left. \psi \right\vert _{t=t_{0}-0}$. The bra/ket product
notation $\left\langle \phi \middle|\psi \right\rangle $ implies integration
of the product $\phi ^{\ast }\psi $ over the interior of box AB.\ For the
potential $V(\mathbf{r})=V(x),$ which depends only on $x$ but not on $y$ and 
$z$ (these are the Cartesian components of the physical coordinate $\mathbf{r%
}$), the eigenstate variables are separated $\Psi _{j}=\tilde{\Psi}%
_{j}(x)\sin (k_{y}y)\sin (k_{z}z)$ so that%
\begin{equation}
-\frac{\hbar ^{2}}{2m}\frac{\partial ^{2}\tilde{\Psi}_{j}}{\partial x^{2}}%
+V(x)\tilde{\Psi}_{j}=\tilde{E}_{j}\tilde{\Psi}_{j}\ \ \ \text{and}\ \ \ 
\tilde{E}_{j}+\hbar ^{2}\frac{k_{y}^{2}+k_{z}^{2}}{2m}=E_{j}  \label{eig1x}
\end{equation}

\subsection{On time-symmetric formulations of quantum mechanics}

The conventional formulation of quantum mechanics implies that the solution $%
\psi $ of the Schr\"{o}dinger equation (\ref{1Shr}) can have only one
temporal boundary condition $\left. \psi \right\vert _{t=t_{0}}=\psi _{0},$
requiring us to set either the initial condition at $t_{0}=-\zeta t_{s}$ or
the final condition at $t_{0}=+\zeta t_{s},$ where $\zeta \geq 1$. Despite
the unitarity and reversibility of quantum evolutions governed by (\ref{1Shr}%
), this violates the symmetry of time and forces us to make a
time-asymmetric choice between the initial and final conditions. Selection
of initial or final conditions is explicitly discriminating in case of
random or diffusional systems \cite{mixing2020}, but one may note that any
given solution $\psi =\psi _{S}(t)$ of the Schr\"{o}dinger equation (\ref%
{1Shr}) allows for two conditions $\left. \psi \right\vert _{t=-t_{s}}=\psi
_{S}(-\zeta t_{s})$ and $\left. \psi \right\vert _{t=+t_{s}}=\psi
_{S}(+\zeta t_{s})$, which correspond to the same $\psi _{S}(t)$. Our
temporal, causality-based intuition, however, forces us to select specific
types of conditions that are associated with antecedent causality and, quite
often, are time-asymmetric. For example, one can choose 1) $\psi =0$ in
section B at $t=-t_{s}$ or 2) $\psi =0$ in section B at $t=+t_{s}$ --- these
conditions are generally not equivalent and, therefore, choosing between
conditions 1 and 2 is time-asymmetric. The fundamental dilemma of selecting
between initial and final conditions is commonly resolved by invoking
antecedent causality and choosing\ initial conditions over final conditions
--- this is practically correct but tends to hide the inequivalence of the
directions of time, especially when some degree of uncertainty is introduced
into the system.

Several interpretation of quantum mechanics permit time-symmetric
formulation of temporal boundary conditions \cite%
{2S-QM1964,TransQM1986,2S-QM2008}. Time reversal is naturally present in
relativistic quantum mechanics due to its \textit{Lorentz invariance}. For \
example, the \textit{Klein--Gordon} (Klein--Gordon--Fock) equation \cite%
{RQM2012}%
\begin{equation}
\frac{1}{c^{2}}\frac{\partial ^{2}\psi }{\partial t^{2}}-\nabla ^{2}\psi +%
\frac{m^{2}c^{2}}{\hbar ^{2}}\psi =0  \label{2-K-G}
\end{equation}%
{\ is of the second order in time, is invariant with respect to the reversal
of time $t\rightarrow -t$ and therefore necessarily involves at least two
waves propagating forward and backward in time. Note that only free spinless
particles satisfy equation (\ref{2-K-G}): interactions of the particle spin
with electromagnetic fields require more elaborate treatment --- the \textit{%
Dirac equation} \cite{Dirac1928}) --- which generally is invariant only
under \textit{charge-parity-time (CPT)} conjugation and not under mere
reversals of time. Interactions of the thermodynamic, time-directional
effects with CPT-invariance have been extensively discussed elsewhere \cite%
{Sakharov1967,Gell-Mann1993,K-PhysA, KM-Entropy2014,SciRep2016,Ent2017} and
are not specifically considered here. The Klein--Gordon equation is used
here only to illustrate the effects of Lorentz invariance.} The
non-relativistic limit of the Klein--Gordon equation is obtained by
substituting $\psi =e^{-i\omega _{0}t}\varphi $ and $\psi =e^{+i\omega
_{0}t}\phi $\ to offset the domination of the $mc^{2}$ term by selecting $%
\omega _{0}=mc^{2}/\hbar $. This yields the two corresponding equations 
\begin{equation}
\text{a) }i\hbar \frac{\partial \varphi }{\partial t}=\mathbb{H}\varphi \ \ 
\text{and b)\ }i\hbar \frac{\partial \phi }{\partial t}=-\mathbb{H}\phi
\label{2rel}
\end{equation}%
where the second equation is the time-reversal $t\rightarrow -t$ of the
first. Conventional non-relativistic quantum mechanics admits only equation (%
\ref{2rel}a), while quantum field theory interprets (\ref{2rel}b) as
corresponding to antiparticles that nominally move backward in time. The 
\textit{transactional interpretation} of quantum mechanics \cite{TransQM1986}
argues that both of these equations play a role: the first corresponds to
waves propagating forward in time and the second corresponds to waves
propagating backward in time and both of these waves are physically
significant.

Another interpretation is given by the so called \textit{two-state vector
formalism} \cite{2S-QM1964,2state1998,2S-QM2008}, where each quantum system
is characterised by two vectors, which are usually written as bra and ket: $%
\left\langle \phi \right\vert $ and $\left\vert \varphi \right\rangle .$
These vectors satisfy the Schr\"{o}dinger equation 
\begin{equation}
\text{a) }i\hbar \frac{\partial \left\vert \varphi \right\rangle }{\partial t%
}=\mathbb{H}\left\vert \varphi \right\rangle \ \ \text{and \ b)\ }i\hbar 
\frac{\partial \left\langle \phi \right\vert }{\partial t}=-\left\langle
\phi \right\vert \mathbb{H}  \label{2st}
\end{equation}%
Equation (\ref{2st}b) can be obtained as the Hermitian (conjugate) transpose
of (\ref{2st}a), although the Hermitian transpose $\varphi ^{\dagger }$of
the state $\varphi $ is not necessarily the same as $\phi ,$ since, as
discussed below, $\left\langle \phi \right\vert $ and $\left\vert \varphi
\right\rangle $ are generally constrained by different initial and final
conditions: the initial conditions are imposed on $\left\vert \varphi
\right\rangle ,$ while $\left\langle \phi \right\vert $ satisfies the final
conditions. {\ Equations (\ref{2st}b) and (\ref{2rel}b) may look similar
but, in fact, these equations are generally different (unless, as in
equation (\ref{1Shr}), the Hamiltonian $\mathbb{H}$ is strictly invariant
with respect to the reversal of time), as are the corresponding conceptual
interpretations. The two-state formalism is conventionally interpreted along
time-asymmetric, casual lines: the state of the system $\left\vert \varphi
\right\rangle $ is determined by its past, while $\left\langle \phi
\right\vert $ specifies how the system will affect measuring devices in the
future, reflecting postselection. According to this casual perspective, $%
\left\vert \varphi \right\rangle $ is a genuine characteristic of the system
at a given moment, while $\left\langle \phi \right\vert $ is not but can be
treated as such for the sake of convenience \cite{2S-QM2008}. There is also
an implied time-symmetric interpretation of the two-state formalism, where
both states $\left\langle \phi \right\vert $ and $\left\vert \varphi
\right\rangle $ are considered to be intrinsic physical characteristics of
the system at a given time moment. }

Finally, the two-state vector formalism requires that the \textit{Born rule}
for the probability density of particle location $P(\mathbf{r)}$, which is
conventionally given by 
\begin{equation}
\left( \overset{\ }{\underset{\ }{P(\text{J,}t)}}\right) _{\text{Born}}=%
\frac{\left\langle \psi \middle|\mathbb{P}_{\text{J}}\middle|\psi
\right\rangle }{Q_{1}}  \label{Born}
\end{equation}%
should be replaced by the time-symmetric \textit{Aharonov, Bergman and
Lebowitz (ABL) rule} \cite{2S-QM1964} 
\begin{equation}
\left( \overset{\ }{\underset{\ }{P(\text{J,}t)}}\right) _{\text{ABL}}=\frac{%
\left\vert \left\langle \phi \middle|\mathbb{P}_{\text{J}}\middle|\varphi
\right\rangle \right\vert ^{2}}{Q_{2}}  \label{ABL}
\end{equation}%
where $\mathbb{P}_{\text{J}},$ J = A,B is projector in the wave function
into either section A or section B and 
\begin{equation}
Q_{1}=\sum_{\text{J}}\left\langle \psi \middle|\mathbb{P}_{\text{J}}\middle%
|\psi \right\rangle ,\ \ \ Q_{2}=\sum_{\text{J}}\left\vert \left\langle \phi %
\middle|\mathbb{P}_{\text{J}}\middle|\varphi \right\rangle \right\vert ^{2}
\end{equation}%
The ABL rule is similar to the interpretation of quantum mechanics called 
\textit{consistent histories} \cite%
{ConsHist1984,Stanford-consistent-histories}. In this context, we stress
that the location operators in (\ref{ABL}) form a consistent set of
projectors since $\mathbb{P}_{\text{A}}\mathbb{P}_{\text{B}}=0$. The
approach of consistent histories also\ has time-symmetric and
time-asymmetric, casual versions of the approach \cite{Gell-Mann1993}.

\subsection{The initial and final conditions}

{\ Leaving aside philosophical aspects of quantum mechanics, we focus on the
initial and final conditions. As specified above, the problem shown in
Figure \ref{fig1} does not have any explicitly measured initial and final
conditions. Such measurements can be performed at $t=-\zeta t_{s}$ and $%
t=+\zeta t_{s}$ with $\zeta >1$ and $\zeta t_{s}\ll t_{b}$, so that the
evolution of the system is not disturbed by these measurements during the
active phase of the experiment. The case of a single particle is discussed
here for the sake of simplicity. The measurements attempt to detect the
presence of particles in section A. If the particle is not detected either
at $t=-\zeta t_{s}$ or at $t=+\zeta t_{s}$ then this realisation is
discarded (i.e. both pre-selection and post-selection apply). The two-state
formalism seems to be the most suitable time-symmetric framework available
for this case. According to this formalism, the state $\left\vert \varphi
\right\rangle $ is deemed to propagate forward in time and, therefore, is
subject to the initial conditions, while $\left\langle \phi \right\vert $ is
deemed to propagate backwards in time and, therefore, is subject to the
final conditions: 
\begin{equation}
\text{a) }\left\vert \varphi \right\rangle _{t=-\zeta t_{s}}=\varphi _{1}%
\text{\ \ \ and \ \ b) }\left\langle \phi \right\vert _{t=+\zeta t_{s}}=\phi
_{2}  \label{2bc}
\end{equation}%
Assuming that the initial and final conditions are the same or similar, so
should be $\varphi _{1}$ and $\phi _{2}$. In conventional quantum mechanics,
we invoke antecedent causality to justify our preference for initial
conditions over final conditions and impose only the initial condition (\ref%
{2bc}a). }

While we can set the initial and final conditions at $t=\pm \zeta t_{s}$,
the corresponding solutions of equations (\ref{2st}) remain undisturbed
until the moments of the potential jumps $t=\pm t_{s}$ are reached. Hence,
from the mathematical perspective, we can put $\zeta =1$ in (\ref{2bc}), and
set these undisturbed conditions at $t=t_{0}=\pm t_{s}$ so that the
equations (\ref{2st}) are to be solved only within the time interval $%
-t_{s}<t<+t_{s}$. Note that, according to the two-state vector formalism,
changes in probabilities may precede the relevant changes of the potential.
The jumps of the potential $V$ in box B at $t_{0}=\pm t_{s}$ are presumed to
be rapid so that the wave functions do not have time to change substantially
and remain practically the same at $t_{0}-0$ and at $t_{0}+0.$ Therefore, we
do not need to specify whether the initial and final conditions are applied
before or after the jumps of the potential. If the final conditions are not
set, the ABL rule (\ref{ABL}) reverts to the Born rule (\ref{Born}).

If the thermalisation time $\tau _{t}$ is smaller than or comparable to $%
\zeta t_{s}$ then, post-selection should have little effect --- the
experiment is effectively screened from the final conditions. If the
characteristic time associated with decoherence $\tau _{d}$ is smaller than
or comparable to $\zeta t_{s}$ (but $\tau _{t}\gg \zeta t_{s}$), then
decoherence can affect the active phase of the experiment by screening it
from the final conditions (in this context decoherence can be seen as an
intermediate projective measurement that remains unknown, i.e. a latent
collapse \cite{Klimenko2019}). If the basis of the measurement and
decoherence are consistent (while accounting for unitary evolution of the
system between the time moments of decoherence and measurement), decoherence
should not have any effects on the measurement. In the opposite case, $\tau
_{d}\gg t_{s}$, decoherence does not have much influence on the experiment
during its active phase $-t_{s}<t<+t_{s}$.

\subsection{Non-intrusive measurements}

If the system under consideration involves a sufficiently large,
statistically significant number of particles $N_{0}$, measurements
conducted over one or few particles should have a minimal effect on the
system. In more accurate terms, this implies that the projection operator $%
\mathbb{P}$ associated with this measurement projects the overall large
Hilbert space into its subspace that has only a slightly smaller dimension
than the original space (i.e. mostly preserving the complexity of the
original state). Measuring interventions, however, involve decoherences and
collapses, and, as demonstrated in experiments \cite{TunnZeno2014}, this can
affect the rate of tunnelling. It is preferable to avoid any irreversible
measurements, at least during the active phase of the experiment. This goal
can be achieved by resorting to generalised or weak measurements \cite%
{WeakM1991,2S-QM2008,WeakMes2013,Stanford-consistent-histories}, which use
an ancilla system.

\begin{figure}[h]
\begin{center}
\includegraphics[width=12cm,page=2,trim=5cm 1cm 2cm 2cm, clip ]{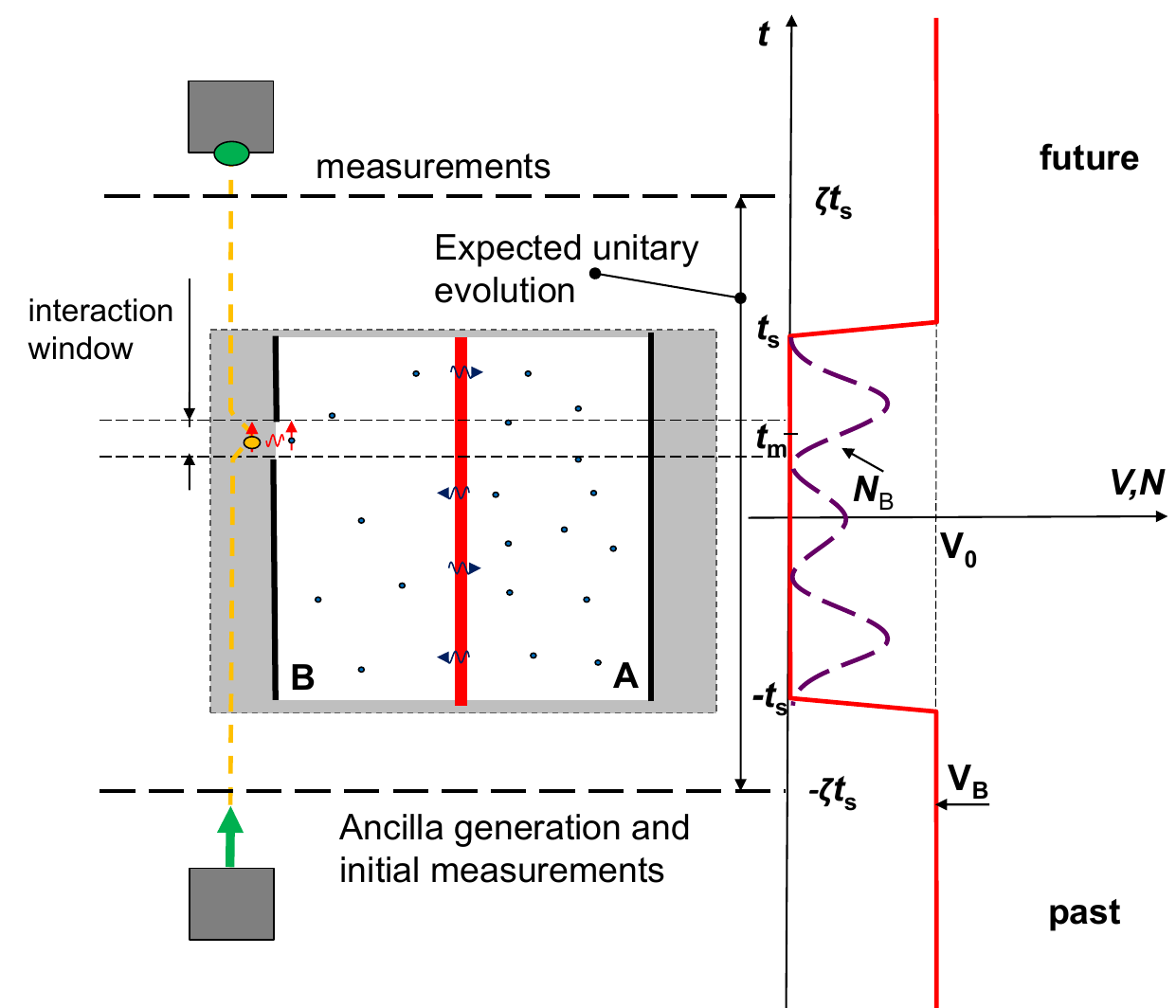}
\caption{Measuring presence of the working particles in section B: the projective
measurements are conducted only on the ancilla system after the active phase of
the experiment is completed. The evolution of the system is unitary during the
active phase. }
\label{fig2}
\end{center}
\end{figure}

The ancilla system is created well before $t=-\zeta t_{s}$ with a few
quantum particles in a specific coherent state, (say, spin down). At some
time moment $t_{m}$ during the active phase $-t_{s}\leq t_{m}\leq +t_{s}$ an
interaction window is created $t_{m}-\Delta t/2\leq t\leq t_{m}+\Delta t/2$
as shown in Figure \ref{fig2}. During this window, unitary interactions are
allowed between the ancilla and tunnelling particles in section B. If
interactions take place, the state of the ancilla particles changes (there
also must be at least some minor change in the state of a particle, say,
alteration of its spin). The state of the ancilla is measured only after $%
t=+\zeta t_{s}$ and alterations of the original ancilla state are indicative
of the presence of tunnelling particles in section B. These type
measurements allow us to detect the presence of tunnelling particles in
section B without causing any decoherence or collapses during the active
phase of the experiment.

\section{Tunnelling without decoherence\label{Sec5}.}

While many tunnelling problems can be solved analytically \cite{LL3,Tunn2003}%
, our goal is in obtaining sufficiently general but relatively simple and
transparent solutions, which are suitable for further analysis involving
decoherence. The initial conditions correspond to all particles located in
section A, presumably in a maximally mixed state although, in this section,
we neglect interactions between the particles and focus on the interaction
of the relevant pure states with the barrier. During the active stage of the
experiment $-t_{s}<t<+t_{s}$, particles tunnel to section B. In this
section, the evolution of quantum particles is examined without the
influence of decoherence so that a coherent wave function remains coherent
during the active phase. For a potential barrier specified by the delta
function $V(x)=s\delta (x),$ we can easily evaluate the energy
eigenfunctions. The probability of tunnelling is presumed to be small $\sim 
\hat{s}^{-2}\ll 1$,\ $\hat{s}=sm/k_{0}\hbar ^{2}.$\ The problem under
consideration involves many possible quantum states but can be effectively
reduced to a two-state dynamic by introducing the partition states. The
resonant, intermediate and non-resonant cases need to be considered
separately. The details of the solutions are elaborated in \ref{SecA}.

\subsection{Evolution of the partition states}

{\ As demonstrated in \ref{SecA}(\ref{sec_near_res}), resonant $\eta
\rightarrow 0,$ near-resonant $\left\vert \eta \right\vert \sim 1$ and
intermediate $1\ll \left\vert \eta \right\vert \ll \hat{s}$ modes form pairs
-- the "plus" mode $\psi _{+}$ and the "minus" mode $\psi _{+}$ with very
close energies and wave numbers. Here, $\eta =2\hat{s}\theta $ and $\theta $
is the phase of the deviation from resonant conditions, i.e. $\theta =0$
corresponds to the exact resonance (see (\ref{A2rm-jAjB}) for details). }
These modes are energy eigenstates and, according to (\ref{1sol}), evolve as 
\begin{equation}
\psi _{\pm }=e^{-i\omega _{\pm }t^{\circ }}\left\vert \pm \right\rangle ,\ \
\ \ \left\vert \pm \right\rangle =\left\{ 
\begin{array}{c}
A_{\pm }\sin (k_{0}x+...)\text{ \ \ in section A} \\ 
B_{\pm }\sin (k_{0}x+...)\text{ \ \ in section B}%
\end{array}%
\right.
\end{equation}%
The conventional normalisation 
\begin{equation}
\frac{x_{\text{{\tiny A}}}\left\vert A_{\pm }\right\vert ^{2}+x_{\text{%
{\tiny B}}}\left\vert B_{\pm }\right\vert ^{2}}{2}=\left\vert \tilde{A}_{\pm
}\right\vert ^{2}+\left\vert \tilde{B}_{\pm }\right\vert ^{2}=1
\end{equation}%
is conveniently expressed in terms of the volume-adjusted amplitudes $\tilde{%
A}$ and $\tilde{B},$ which satisfy 
\begin{equation}
\left( \frac{\tilde{B}}{\tilde{A}}\right) _{-}\left( \frac{\tilde{B}}{\tilde{%
A}}\right) _{+}=-1,\ \ \tilde{A}=A\sqrt{\frac{x_{\text{{\tiny A}}}}{2}},\ \ 
\tilde{B}=B\sqrt{\frac{x_{\text{{\tiny B}}}}{2}}  \label{A5rat}
\end{equation}%
according to (\ref{A2rm-rat}).

The stationary orthogonal (unitary) transformation of the basis 
\begin{eqnarray}
\left\vert \text{A}\right\rangle &=&\frac{1}{\sqrt{1+\xi ^{2}}}\left( \ \big|%
+\big\rangle\ +\ \xi \big|-\big\rangle\right)  \label{A5A} \\
\left\vert \text{B}\right\rangle &=&\frac{1}{\sqrt{1+\xi ^{2}}}\left( \xi %
\big|+\big\rangle\ -\ \ \big|-\big\rangle\right)  \label{A5B}
\end{eqnarray}%
converts the "plus" $\left\vert +\right\rangle $ and "minus" $\left\vert
-\right\rangle $ eigenstates, into states $|$A$\rangle $ and $|$B$\rangle $.
Unlike the states $\left\vert +\right\rangle $ and $\left\vert
-\right\rangle ,$ the states $|$A$\rangle $ and $|$B$\rangle $ are not
energy eigenstates. Here we denote%
\begin{equation}
\xi =\left( \frac{\tilde{B}}{\tilde{A}}\right) _{+}=\sqrt{\frac{x_{\text{%
{\tiny B}}}}{x_{\text{{\tiny A}}}}}\left( \frac{B}{A}\right) _{+}=\sigma
\left( \frac{x_{\text{{\tiny A}}}}{x_{\text{{\tiny B}}}}\right)
^{1/2}F_{+}\left( \eta ,\frac{x_{\text{{\tiny B}}}}{x_{\text{{\tiny A}}}}%
\right)  \label{A5AB}
\end{equation}%
where the function $F_{+}$ is defined by (\ref{A2rm-BA}) and $\sigma =\pm 1$
as specified in (\ref{A2rm-eq2}). The states $|$A$\rangle $ and $|$B$\rangle 
$ are referred to as the "partition states": to the leading order of our
analysis, the state $|$A$\rangle $ implies exclusive localisation in section
A of the box, while the state $|$B$\rangle $ corresponds to exclusive
localisation in section B. Indeed, with the definition of $\xi $ given by (%
\ref{A5AB}) and the use of equations (\ref{3PSI}), (\ref{A2rm-jAjB})-(\ref%
{A2rm-rat}) the partition states are approximated by 
\begin{equation}
\left\vert \text{A}\right\rangle \approx \left\{ 
\begin{array}{cc}
\left( \frac{2}{x_{\text{{\tiny A}}}}\right) ^{1/2}\sin (k_{0}x+...) & \text{%
in section A} \\ 
0 & \text{in section B}%
\end{array}%
\right.
\end{equation}%
\begin{equation}
\left\vert \text{B}\right\rangle \approx \left\{ 
\begin{array}{cc}
0 & \text{in section A} \\ 
\left( \frac{2}{x_{\text{{\tiny B}}}}\right) ^{1/2}\sin (k_{0}x+...) & \text{%
in section B}%
\end{array}%
\right.
\end{equation}%
since $k_{+}=\Delta k_{+}+k_{0}\approx k_{-}=\Delta k_{-}+k_{0}\approx k_{0}$
at the leading order and we select $A_{+}>0$ and $A_{-}>0$ to remove freedom
in choosing signs.

It is clear that the normalised amplitudes of the wave functions that
correspond to states $|$A$\rangle $ and $|$B$\rangle $\ are given by $\tilde{%
A}$ and $\tilde{B}$. Since the states $|$A$\rangle $ and $|$B$\rangle $\ are
not energy eigenstates, their amplitudes $\tilde{A}$ and $\tilde{B}$ change
in time as determined by the equation%
\begin{equation}
i\hbar \frac{\partial }{\partial t}\left[ 
\begin{array}{c}
\tilde{A} \\ 
\tilde{B}%
\end{array}%
\right] =\mathbb{H}\left[ 
\begin{array}{c}
\tilde{A} \\ 
\tilde{B}%
\end{array}%
\right] ,\ \ \mathbb{H}=\mathbb{H}_{0}+\mathbb{H}_{1},\ \   \label{U_AB}
\end{equation}%
where the Hamiltonian in the new basis is given by%
\begin{equation}
\mathbb{H}_{0}=\left[ 
\begin{array}{cc}
\frac{E_{+}+E_{-}}{2} & 0 \\ 
0 & \frac{E_{+}+E_{-}}{2}%
\end{array}%
\right] ,\ \ \mathbb{H}_{1}=\frac{E_{+}-E_{-}}{1+\xi ^{2}}\left[ 
\begin{array}{cc}
\frac{\left( 1-\xi ^{2}\right) }{2} & \xi \\ 
\xi & \frac{\left( \xi ^{2}-1\right) }{2}%
\end{array}%
\right]  \label{H0H1}
\end{equation}%
Here, $\left\langle +\right\vert \mathbb{H}\left\vert +\right\rangle =E_{+}$%
, $\left\langle -\right\vert \mathbb{H}\left\vert -\right\rangle =E_{-}$ and 
$\left\langle +\right\vert \mathbb{H}\left\vert -\right\rangle =\left\langle
-\right\vert \mathbb{H}\left\vert +\right\rangle =0$ since the states $%
\left\vert +\right\rangle $ and $\left\vert -\right\rangle $ are energy
eigenstates. According to (\ref{1sol}), this equation is solved by the
following unitary evolution matrix $\mathbb{U}$%
\begin{equation}
\left[ 
\begin{array}{c}
\tilde{A} \\ 
\tilde{B}%
\end{array}%
\right] =\underset{\mathbb{U}}{\underbrace{\frac{\Omega _{0}}{1+\xi ^{2}}%
\left[ 
\begin{array}{cc}
\Omega +\xi ^{2}/\Omega & -2i\xi \sin \left( \frac{\Delta \omega }{2}%
t^{\circ }\right) \\ 
-2i\xi \sin \left( \frac{\Delta \omega }{2}t^{\circ }\right) & \xi
^{2}\Omega +1/\Omega%
\end{array}%
\right] }}\left[ 
\begin{array}{c}
\tilde{A} \\ 
\tilde{B}%
\end{array}%
\right] _{t=t_{0}}  \label{A5U}
\end{equation}%
where we denote%
\begin{equation}
E_{0}=\frac{E_{+}+E_{-}}{2}\approx \frac{k_{0}^{2}\hbar ^{2}}{2m},\text{\ \ }%
\Omega _{0}=e^{-\frac{iE_{0}t^{\circ }}{\hbar }},\ \text{\ }\Omega =e^{-i%
\frac{\Delta \omega }{2}t^{\circ }},\ \ t^{\circ }=t-t_{0},  \label{A5eq1}
\end{equation}%
\begin{equation}
\Delta \omega =\omega _{+}-\omega _{-}=\frac{E_{+}-E_{-}}{\hbar }\approx 
\frac{k_{0}\hbar }{m}\Delta k,\ \ \ \Delta k=\Delta k_{+}-\Delta k_{-}=\frac{%
1}{2\hat{s}}\frac{D^{1/2}}{x_{\text{{\tiny A}}}x_{\text{{\tiny B}}}}
\label{A5eq2}
\end{equation}%
and $D$ is specified in (\ref{A2rm-D}). With the initial conditions 
\begin{equation}
\left[ 
\begin{array}{c}
\tilde{A} \\ 
\tilde{B}%
\end{array}%
\right] _{t=t_{0}}=\left[ 
\begin{array}{c}
1 \\ 
0%
\end{array}%
\right]
\end{equation}%
which correspond to particle location in section A at $t=t_{0}$, the
amplitudes of the partition states depend on time $t^{\circ }=t-t_{0}$ and
evolve as 
\begin{equation}
\left[ 
\begin{array}{c}
\tilde{A} \\ 
\tilde{B}%
\end{array}%
\right] =\frac{\Omega _{0}}{1+\xi ^{2}}\left[ 
\begin{array}{c}
e^{-i\frac{\Delta \omega }{2}t^{\circ }}+\xi ^{2}e^{+i\frac{\Delta \omega }{2%
}t^{\circ }} \\ 
-2i\xi \sin \left( \frac{\Delta \omega }{2}t^{\circ }\right)%
\end{array}%
\right]  \label{ABt}
\end{equation}%
assuming that all particles are initially present only in the section A.
Note that the evolution preserves normalisation $|\tilde{A}|^{2}+|\tilde{B}%
|^{2}=1,$ where the amplitudes $|\tilde{A}|^{2}$ and $|\tilde{B}|^{2}$ are
conventionally interpreted as probabilities of localisation $P($A$)=|\tilde{A%
}|^{2}$ and $P($B$)=|\tilde{B}|^{2}$ associated with this resonant pair. The
extent of tunnelling (i.e. a quantity constraining $|\tilde{B}|$ and $P($B$)$
for any $t^{\circ }$), which is determined by $\varsigma =\left\vert \xi
\right\vert /(1+\xi ^{2})$, remains small when $\left\vert \xi \right\vert
\ll 1$ or $\left\vert \xi \right\vert \gg 1$.

\subsection{Tunnelling by resonant and near-resonant modes}

The resonant modes ($\eta \rightarrow 0$) are energy eigenstates that are
energy eigenstates in all sections of the box, that is resonant modes are
resonant in both sections A and B. The near-resonant modes are ($\left\vert
\eta \right\vert \sim 1$) close to the resonant conditions in A and B. For
these modes, the characteristic transmission frequency $\hat{\omega}_{\text{r%
}}$ and the characteristic transmission time $\hat{\tau}_{\text{r}}=1/\hat{%
\omega}_{\text{r}}$ are evaluated from equations (\ref{A5AB}), (\ref{A5eq1})
and\ (\ref{A5eq2}) 
\begin{equation}
\xi =\sigma \sqrt{\frac{x_{\text{{\tiny A}}}}{x_{\text{{\tiny B}}}}}%
F_{+}\left( \eta ,\frac{x_{\text{{\tiny B}}}}{x_{\text{{\tiny A}}}}\right)
,\ \ \ \Delta \omega =\frac{u_{0}}{2\hat{s}}\frac{D^{1/2}}{x_{\text{{\tiny A}%
}}x_{\text{{\tiny B}}}}  \label{taunr}
\end{equation}%
where $F_{+}$ and $D$ depend on $\eta =2\hat{s}\theta ,$ $x_{\text{{\tiny A}}%
}$ and $x_{\text{{\tiny B}}}$\ as specified in (\ref{A2rm-dk}) -- (\ref%
{A2rm-D}). For the resonance modes $\eta \rightarrow 0,$ these equations
simplify according to (\ref{A4dk}) and (\ref{A4BA}): 
\begin{equation}
\xi =\sigma \sqrt{\frac{x_{\text{{\tiny B}}}}{x_{\text{{\tiny A}}}}}\text{,\
\ \ }\Delta \omega =\hat{\omega}_{\text{r}}=\frac{1}{\hat{\tau}_{\text{r}}}%
\approx \frac{u_{0}}{x_{0}\hat{s}}=\frac{1}{\tau _{0}\hat{s}}  \label{taur}
\end{equation}%
Here, we also introduce useful parameters$\ $%
\begin{equation}
\tau _{0}=\frac{x_{0}}{u_{0}},\ u_{0}=\frac{k_{0}\hbar }{m},\ \ x_{0}=\frac{%
2x_{\text{{\tiny A}}}x_{\text{{\tiny B}}}}{x_{\text{{\tiny A}}}+x_{\text{%
{\tiny B}}}}
\end{equation}%
\ where $\tau _{0}$ is the characteristic fly time defined in terms of the
characteristic length of the box section $x_{0}$ and the characteristic
velocity $u_{0},$ which can be estimated using thermodynamic quantities $%
mu_{0}^{2}=2\tilde{E}\approx k_{B}T$.

If $x_{\text{{\tiny B}}}=x_{\text{{\tiny A}}}=x_{0},$ then all modes are
resonant, while the "minus" mode becomes symmetric and the "plus" mode
antisymmetric --- this case is referred to as the resonance case (see \ref%
{SecA}(\ref{sec_res})). In the resonance case, the evolution of the
partition states simplifies into 
\begin{equation}
\left[ 
\begin{array}{c}
\tilde{A} \\ 
\tilde{B}%
\end{array}%
\right] =\Omega _{0}\left[ 
\begin{array}{c}
+\cos \left( \frac{\hat{\omega}_{\text{r}}}{2}t^{\circ }\right) \\ 
-i\sin \left( \frac{\hat{\omega}_{\text{r}}}{2}t^{\circ }\right)%
\end{array}%
\right]
\end{equation}

\subsection{Tunnelling by intermediate and non-resonant modes}

We now examine the limit $\eta \rightarrow \pm \infty $ and turn to
consideration of the intermediate ($1\ll \left\vert \eta \right\vert \ll 
\hat{s}$) and non-resonant modes ($\left\vert \eta \right\vert \sim \hat{s}$%
), which as shown in \ref{SecA}(\ref{sec_non_res}) must be either A-resonant
or B-resonant, assuming that $\hat{s}\gg 1$. Generally, links between the
plus and minus modes are preserved for the intermediate modes but weaken for
non-resonant modes, which do not necessarily form pairs. \ref{SecA}(\ref%
{sec_non_res}) indicates that $\left\vert B/A\right\vert \sim 1/\left\vert
\eta \right\vert \ll 1$ for A-resonant modes and $\left\vert B/A\right\vert
\sim \left\vert \eta \right\vert \gg 1$ for B-resonant modes. Since the
initial wave function $\psi _{0}=\left. \psi \right\vert _{t=t_{0}}$ is
localised exclusively in section A$,$\ the A-resonant modes dominate the
expansion in energy eigenstates (\ref{1sol})-(\ref{eig1r}). The components
with different values of $k$ and $\omega $ quickly lose phase correlation
and we focus on modes that have close $k$ and $\omega $. If $\eta
\rightarrow +\infty ,$ the "plus" branch corresponds to A-resonant modes and
the "minus" branch corresponds to B-resonant modes. Equations (\ref{A5A}), (%
\ref{A5B}) and (\ref{ABt}) are still valid for intermediate modes, but the
solution parameters are evaluated differently 
\begin{equation}
\xi =\sigma \frac{\sqrt{x_{\text{{\tiny B}}}x_{\text{{\tiny A}}}}}{x_{\text{%
{\tiny A}}}+x_{\text{{\tiny B}}}}\frac{1}{\eta },\ \Delta k\approx \frac{%
\left\vert \eta \right\vert }{\hat{s}}\frac{1}{x_{0}},\ \ \ \Delta \omega =%
\hat{\omega}_{\text{i}}=\frac{1}{\hat{\tau}_{\text{i}}}\sim \frac{\left\vert
\eta \right\vert }{\hat{s}}\frac{u_{0}}{x_{0}}  \label{IR-1}
\end{equation}%
from equations (\ref{A5eq1}),\ (\ref{A5eq2}), (\ref{A2dk_inf}) and (\ref%
{A2BA_inf}) assuming $\eta =2\hat{s}\theta \rightarrow \pm \infty $. For the
non-resonant modes, we can use the same estimates but put $\left\vert \theta
\right\vert \approx 1$, $\left\vert \eta \right\vert \sim \hat{s}$ and
estimate 
\begin{equation}
\left\vert \xi \right\vert \sim \frac{1}{\hat{s}}\ll 1,\ \Delta k\sim \frac{1%
}{x_{0}},\ \ \ \Delta \omega =\hat{\omega}_{\text{n}}=\frac{1}{\hat{\tau}_{%
\text{n}}}\sim \frac{u_{0}}{x_{0}}=\frac{1}{\tau _{0}}  \label{NR-1}
\end{equation}%
The modes away from the resonance conditions are characterised by relatively
small extent of tunnelling determined by $\varsigma \approx \xi $. The
probability of localisation in section B delivered by the intermediate modes
evolves periodically 
\begin{equation}
P(\text{B},t)=\left\vert \tilde{B}\right\vert ^{2}\approx 4\xi ^{2}\sin
^{2}\left( \frac{\Delta \omega }{2}\left( t-t_{0}\right) \right) \sim \frac{1%
}{\eta ^{2}}\ll 1  \label{NR-2}
\end{equation}%
and becomes small $\sim 1/\hat{s}^{2}$ for non-resonant modes (despite
progressing faster in time than the resonant modes $\left\vert
t-t_{0}\right\vert \sim \hat{\tau}_{\text{n}}\sim \tau _{0}\ll \hat{\tau}_{%
\text{r}}$). Note that the resonant modes also achieve probability $P($B$%
,t)\sim 1/\hat{s}^{2}$ over time $t\sim \tau _{0}$ but, unlike the
non-resonant modes, they proceed further to deliver $P($B$,t)\sim 1$ when $%
t\sim \hat{\tau}_{\text{r}}\sim \tau _{0}\hat{s}\gg \tau _{0}$.

The estimates of this section (\ref{NR-1}) and (\ref{NR-2}) remain the same
even if a non-resonant mode (say, A-resonant but not B-resonant) is not
explicitly coupled with any B-resonant mode. Indeed, over time $t\sim \tau
_{n}\sim \tau _{0},$ this mode would lose correlations with the\ other modes
and according to (\ref{A2Ar}) can contribute to particles appearing in
section B only a small probability $\sim 1/\hat{s}^{2}$ at any time $%
t\gtrsim \hat{\tau}_{\text{n}}$. Without a sufficiently large fraction of
the resonant modes, the probability of finding tunnelling particles in
section B remains small indefinitely. While tunnelling is contributed less
by non-resonant modes, these modes are often more numerous than the resonant
and intermediate modes.

The fraction of resonant modes is determined by geometry, i.e. by $x_{\text{%
{\tiny A}}}$ and $x_{\text{{\tiny B}}}$. For example, if $x_{\text{{\tiny A}}%
}=2x_{\text{{\tiny B}}},$ then each second A-resonant mode is also
B-resonant. Note that under conditions of $x_{\text{{\tiny A}}}\sim x_{\text{%
{\tiny B}}}$ the fraction of resonant and near-resonant modes cannot fall
below $\sim 1/\hat{s}$. First, let us assume $x_{\text{{\tiny A}}}\geq x_{%
\text{{\tiny B}}}$ or, otherwise, swap A and B. \ref{SecA}(\ref{sec_near_res}%
) indicates that $\theta \sim 1/\hat{s}$ to achieve near-resonance
conditions. Equations (\ref{A2rm-jAjB}) result in $j_{\text{{\tiny B}}%
}=\gamma j_{\text{{\tiny A}}}-\theta ^{\prime }$ where $j_{\text{{\tiny B}}}$
and $j_{\text{{\tiny A}}}$ are integers, $\gamma =$ $x_{\text{{\tiny B}}}/x_{%
\text{{\tiny A}}}\leq 1$, and $\theta ^{\prime }=\theta (1+x_{\text{{\tiny B}%
}}/x_{\text{{\tiny A}}})/\pi \sim 1/\hat{s}$ is small. Assuming that $j_{%
\text{{\tiny B}}}$ can reach $1$ for typical energies, the overall fraction
of resonant and near-resonant modes $\left\vert j_{\text{{\tiny B}}}-\gamma
j_{\text{{\tiny A}}}\right\vert \lesssim 1/\hat{s}$ of all modes $j_{\text{%
{\tiny A}}}=1,2,3,...$ and $j_{\text{{\tiny B}}}=1,2,3,...$, cannot be
smaller than $\sim 1/\hat{s}$. Therefore, despite being relatively small in
the numbers of modes, these numbers are sufficient for the resonant and
near-resonant modes to dominate tunnelling.

\section{Effect of decoherence on tunnelling\label{Sec6}}

This section examines the effects of decoherence on tunnelling, which appear
to be substantial and, therefore, detectable in experiments. We begin with a
general consideration of decoherence leading to a specific form of the 
\textit{Lindblad equation} \cite{LindbladG1976} that corresponds to our
understanding of decoherence. The exact physical mechanism responsible for
decoherence remains unknown and referred to here as time priming; while the
main parameter that quantifies decoherence is its characteristic frequency $%
\omega _{d}$. Decoherence is expected to result in loss of coherent
interferences without any substantial unitary interactions with the
environment \cite{ImryYoseph2002,Dattagupta2004}, measurements \cite%
{TunnZeno2014}, or any other effects that may cause a significant
redistribution of energy. With exception of the last subsection, we consider
effects that are intrinsic or effectively intrinsic. While decoherence
triggers equilibration and thermalisation, the latter does involve a
redistribution of energy and should not be confused with decoherence, which
is deemed to have negligible energy effects. The obtained form of the
Lindblad equation is converted into a \textit{Pauli master equation} \cite%
{Pauli1928} for state probabilities, which is subsequently used for
determining the effect of decoherence on the tunnelling rates\ in the
resonant and nonresonant cases.

\subsection{Decoherence in the context of time priming.}

While unitary evolution of a quantum system is fully specified by the Schr%
\"{o}dinger equation, our knowledge of decoherence and collapses is much
more limited. Let us illustrate this point by a simple example: consider two
states of a quantum system 
\begin{equation}
\psi _{+}=\frac{1}{\sqrt{2}}\left( \left\vert E_{1}\right\rangle +\left\vert
E_{2}\right\rangle \right) \text{ and }\psi _{-}=\frac{1}{\sqrt{2}}\left(
\left\vert E_{1}\right\rangle -\left\vert E_{2}\right\rangle \right)
\end{equation}%
that are expressed in terms the energy eigenstates $\left\vert
E_{1}\right\rangle $ and $\left\vert E_{2}\right\rangle $. The superposition
state $\psi =(\psi _{+}+\psi _{-})/2^{1/2}=\left\vert E_{1}\right\rangle $
would have its energy measured as $E_{1}$. Assume that $\psi $ decoheres
into a mixture of $\psi _{+}$ and $\psi _{-}$ with equal probabilities.
Measuring energy for each of these functions $\psi _{+}$ and $\psi _{-}$
would produce either $E_{1}$ or $E_{2}$ with equal probability. The choice
of $\psi _{+}$ and $\psi _{-}$ as the decoherence basis that does not
coincide with energy eigenstates results in substantial energy change in the
system. In the context of the direction of time, however, decoherence is
commonly understood as loss of interference between the components with
minimal energy interactions. This, of course, does not exclude other forms
of decoherence with stronger interactions and significant energy exchanges
and these other forms may be important under some conditions. In the present
work, however, we restrict our attention to less energetic forms of
decoherence that can be associated with the time primer.

Still, choosing exact energy eigenstates as the basis for decoherence does
not solve the problem --- these eigenstates continue to exist without
interacting with each other and this is not a particularly interesting case.
Analysis of decoherence becomes most meaningful when decoherence basis is
selected along with eigenstates of the principal part $\mathbb{H}_{0}$ of
the Hamiltonian $\mathbb{H=H}_{0}+\mathbb{H}^{\prime }$ but there is also a
smaller interference component $\mathbb{H}^{\prime }$ that acts along with
decoherence. It is clear that splitting the Hamiltonian in two parts
requires some physical grounds for doing this. For example $\mathbb{H}_{0}$
may be Hamiltonian that is intrinsically associated with a system, while $%
\mathbb{H}^{\prime }$ corresponds to external influence or some other form
of interference. In context of particle physics, $\mathbb{H}_{0}$ is
conventionally related to strong interactions, while $\mathbb{H}^{\prime }$
pertains to weak interactions that are known to break the symmetry of the
directions of time in CP violations (which seems to be consistent with
time-directional character of decoherence). In any case, the decoherence
basis that is associated with eigenstates of some Hamiltonian $\mathbb{H}%
_{0} $ must be orthogonal (and is conventionally selected orthonormal). The
wave functions undergo unitary transformations $\left\vert \psi
\right\rangle _{t^{\prime }}=\mathbb{U}(t^{\prime }-t)\left\vert \psi
\right\rangle _{t}$ but may also experience decoherence events, where the
projections $\left\vert d_{j}\right\rangle \left\langle d_{j}\right\vert
\left\vert \psi \right\rangle $ of every wave function $\psi $ onto the
decoherence basis $\left\vert d_{1}\right\rangle ,...,\left\vert
d_{n}\right\rangle $ lose their coherence (completely or partially). As
considered above, $\left\langle d_{i}\right\vert \left\vert
d_{j}\right\rangle =\delta _{ij}$ since $\left\vert d_{j}\right\rangle $
satisfies $\mathbb{H}_{0}\left\vert d_{j}\right\rangle =E_{j}^{\circ
}\left\vert d_{j}\right\rangle $.

Consider the density matrix $\mathbf{\rho ,}$ which generally evolves by
unitary transformations $\mathbf{\rho }^{\prime }=\mathbb{U}\mathbf{\rho }%
\mathbb{U}^{\dagger }$ but also experiences\ decoherence events, where it is
transformed by the Kraus operators $\mathbb{K}_{j}$%
\begin{equation}
\mathbf{\rho }^{\prime }\mathbf{=}\sum_{j=0}^{n}\mathbb{K}_{j}\mathbf{\rho }%
\mathbb{K}_{j}^{\dagger },\ \ \ \ \ \ \sum_{j=0}^{n}\mathbb{K}_{j}^{\dagger }%
\mathbb{K}_{j}=\mathbb{I}\   \label{Kraus}
\end{equation}%
Note that equation (\ref{Kraus}) represents a specific form of the \textit{%
Kraus transformation} 
\begin{equation}
\mathbb{K}_{j}=\sqrt{\lambda }\left\vert d_{j}\right\rangle \left\langle
d_{j}\right\vert ,\ \ \ \mathbb{K}_{0}=\sqrt{(1-\lambda )}\mathbb{I}
\end{equation}%
that corresponds to specific action of decoherence that is discussed above
(assuming $0\leq \lambda \leq 1$) and, at least in principle, can be
associated with the time primer. The last constraint in (\ref{Kraus}) is
satisfied since, obviously, $\Sigma _{j}\left\vert d_{j}\right\rangle
\left\langle d_{j}\right\vert =\mathbb{I}$. If $\lambda =0$, transformation (%
\ref{Kraus}) is identical $\mathbf{\rho }^{\prime }\mathbf{=}\mathbb{I}%
\mathbf{\rho }$ and no decoherence occurs. If $\lambda =1$, the projections
on the decoherence basis becomes fully independent. The transformation (\ref%
{Kraus}) with $\lambda ^{\prime }=1-(1-\lambda )^{-1}$ reverses (\ref{Kraus}%
) with $\lambda $ and, therefore, represents recoherence. While Kraus
operators are convenient to characterise discrete decoherence events that
interrupt unitary evolution, continuous decoherence can be conventionally
described by the Lindblad operators $\mathbb{L}_{j}=\mathbb{K}_{j}/\sqrt{%
\lambda }$ so that 
\begin{equation}
\mathbf{\rho }^{\prime }\mathbf{=}(1-\lambda )\mathbf{\rho }+\lambda
\sum_{j=1}^{n}\mathbb{L}_{j}\mathbf{\rho }\mathbb{L}_{j}^{\dagger }
\end{equation}%
Assuming that $\lambda =\Delta t/\tau _{d}$, we obtain 
\begin{equation}
\Delta \mathbf{\rho }=\mathbf{\rho }^{\prime }-\mathbf{\rho =}\frac{\Delta t%
}{i\hbar }\left[ \mathbb{H},\mathbf{\rho }\right] +\frac{\Delta t}{\tau _{d}}%
\left( \sum_{j=1}^{n}\mathbb{L}_{j}\mathbf{\rho }\mathbb{L}_{j}^{\dagger }-%
\mathbf{\rho }\right)  \label{Lin1}
\end{equation}%
where the Hamiltonian term reflects a time differential of the unitary
transformation $\mathbf{\rho }^{\prime }=\mathbb{U}\mathbf{\rho }\mathbb{U}%
^{\dagger }.$ Dividing this equation by $\Delta t$ and taking the limit $%
\Delta t\rightarrow 0$ leads to a specific, simple form of the \textit{%
Lindblad equation} 
\begin{equation}
\frac{\partial \mathbf{\rho }}{\partial t}=\frac{\left[ \mathbb{H},\mathbf{%
\rho }\right] }{i\hbar }+\frac{1}{\tau _{d}}\left( \sum_{j=1}^{n}\mathbb{L}%
_{j}\mathbf{\rho }\mathbb{L}_{j}^{\dagger }-\mathbf{\rho }\right)
\label{Lin2}
\end{equation}%
The simplification of the Lindblad equation is due to the relation $\tsum_{j}%
\mathbb{L}_{j}^{\dagger }\mathbb{L}_{j}=\mathbb{I}$, which is valid here but
not satisfied in the general case. The value $\tau _{d}$ represents the
characteristic decoherence time. Since $\mathbb{L}_{j}$ are Hermitian and $%
\tau _{d}>0$ in this form of the Lindblad equation, the evolution governed
by (\ref{Lin2}) does not decrease entropy \cite{Abe2017}.

While the physical implications of discrete and continuous decoherence
should be similar, we, for the sake of transparency, consider discrete
decoherence events specified by (\ref{Kraus}) with $\lambda =1$ and spaced
by characteristic decoherence time $\tau _{d}$. The decoherence events
suppress all non-diagonal elements of the density matrix, while the unitary
evolution $\mathbf{\rho }^{\prime }\mathbf{=}\mathbb{U}\mathbf{\rho }\mathbb{%
U}^{\dagger }$ persists between the decoherence\ events \cite{SciRep2016}.
Hence, the density matrix is transformed by the unitary evolution and a
subsequent decoherence event as 
\begin{equation*}
\left[ 
\begin{array}{ccc}
\rho _{11} &  & 0 \\ 
& \ddots &  \\ 
0 &  & \rho _{nn}%
\end{array}%
\right] \underset{t\rightarrow t^{\prime }}{\longrightarrow }\left[ 
\begin{array}{ccc}
\rho _{11}^{\prime } &  & 0 \\ 
& \ddots &  \\ 
0 &  & \rho _{nn}^{\prime }%
\end{array}%
\right] ,\ \ \ \rho _{jj}^{\prime }=\sum_{k}\left\vert U_{kj}\right\vert
^{2}\rho _{kk}
\end{equation*}%
over each of the intervals $[t,t^{\prime }],$ where $t^{\prime }=t+\tau _{d}$%
. Here, $U_{kj}$ represent the components of the unitary evolution operator $%
\mathbb{U}(t^{\prime }-t)$. Considering long times $t>\tau _{d}$, we
conclude that the probabilities $P_{j}=\rho _{jj}$ are transformed according
to 
\begin{equation}
\frac{dP_{j}}{dt}=\frac{1}{\tau _{d}}\sum_{k}\left( \left\vert
U_{kj}\right\vert ^{2}-\delta _{kj}\right) P_{k}  \label{PauliME}
\end{equation}%
which, essentially, is a \textit{Pauli master equation} describing evolution
of a Markov chain with transitional probabilities given by deviations of $%
\left\vert U_{kj}\right\vert ^{2}$ from the unity matrix.

Despite the existence of many theories \cite%
{Joos2003,Schlosshauer2007,Zeh2007}, there is no certainty about the exact
effect of decoherence on wave functions distributed in space. We, however,
expect loss of coherence between energy eigenstates with substantially
different energies$,$ as well as expect and are primarily interested in
losses of coherence between the branches of the wave functions located in
sections A and B, which converts coherent waves into a mixture of
probabilities for particle presence in these sections. In any case,
decoherence can be charactetised by its principal parameter --- the
characteristic frequency of decoherence $\omega _{d}$\ or the characteristic
decoherence time $\tau _{d}=1/\omega _{d}$, which is featured in equation (%
\ref{PauliME}).

\subsection{Effect on the resonant and near-resonant modes}

If the characteristic time of decoherence is longer than the resonance
tunnelling time $\tau _{d}>\hat{\tau}_{\text{r}}\approx \tau _{0}\hat{s},$
decoherence has little effect on the tunnelling rate but even infrequent
decoherence changes the character of the solution --- it relaxes towards
stationary distributions instead of oscillating indefinitely. When, however,
the decoherence time becomes short $\tau _{d}<\hat{\tau}_{\text{r}}$ (but
not too short $\tau _{d}>\tau _{0}$) it converts the unitary evolution of
the resonance modes, which is specified by (\ref{U_AB})-(\ref{A5U}) and (\ref%
{ABt}) in the basis of the partition states, into a Markov process, which
according to (\ref{PauliME}) is given by%
\begin{equation}
\frac{d}{dt}\left[ 
\begin{array}{c}
P_{\text{{\tiny A}}} \\ 
P_{\text{{\tiny B}}}%
\end{array}%
\right] =\frac{1}{\tau _{d}}\left[ 
\begin{array}{cc}
-W & W \\ 
W & -W%
\end{array}%
\right] \left[ 
\begin{array}{c}
P_{\text{{\tiny A}}} \\ 
P_{\text{{\tiny B}}}%
\end{array}%
\right] =\frac{1}{\tilde{\tau}_{\text{r}}}\left[ 
\begin{array}{cc}
-1 & 1 \\ 
1 & -1%
\end{array}%
\right] \left[ 
\begin{array}{c}
P_{\text{{\tiny A}}} \\ 
P_{\text{{\tiny B}}}%
\end{array}%
\right]  \label{Markov2}
\end{equation}%
where $\tilde{\tau}_{\text{r}}=W/\tau _{d},$ $W=\left\vert U_{\text{{\tiny AB%
}}}\right\vert ^{2}\ $and $\left\vert U_{\text{{\tiny AB}}}\right\vert \sim
2\varsigma \sin \left( \Delta \omega \tau _{d}/2\right) ,\ \ \varsigma
=\left\vert \xi \right\vert /(1+\xi ^{2})$\ is the off-diagonal component of
the unitary evolution matrix (\ref{A5U}) and $\Delta \omega $ is given by (%
\ref{taunr}). Note that, according to (\ref{Markov2}), the extent of
tunnelling is no longer limited by $\varsigma$. Since $\Delta \omega
^{-1}\sim \hat{\tau}_{\text{r}}>\tau _{d}$, the sine can be expanded $%
\left\vert U_{\text{{\tiny AB}}}\right\vert \sim \varsigma \Delta \omega
\tau _{d}$. For exact resonance, we assume $x_{\text{{\tiny B}}}\sim x_{%
\text{{\tiny A}}}$, put $\eta =0$, $\varsigma \sim 1$ \ and obtain 
\begin{equation}
\tilde{\omega}=\frac{1}{\tilde{\tau}}\approx \hat{\omega}_{\text{r}}^{2}\tau
_{d}=\frac{\tau _{d}}{\hat{\tau}_{\text{r}}^{2}}=\frac{\tau _{d}}{\tau
_{0}^{2}\hat{s}^{2}}  \label{w-r-d}
\end{equation}%
This expression reflects the quantum Zeno effect, which is well-known and
has been recently demonstrated in experiments \cite{TunnZeno2014} applying
frequent measurements to quantum tunnelling. Increasing frequency of
decoherence reduces the rate of tunnelling for resonant modes. Note that,
even in the presence of decoherence, the resonant modes do not lead to the
same density of particles in both sections (when $x_{\text{{\tiny A}}}\neq
x_{\text{{\tiny B}}}$) but to the equal probabilities of being in these
sections $P_{\text{{\tiny A}}},P_{\text{{\tiny B}}}\rightarrow 1/2$ as $%
t\rightarrow \infty $. This is consistent with general expectations of
statistical quantum mechanics: the amplitudes of modes having similar
energies are expected to be similar under equilibrium conditions.

\subsection{Effect on the non-resonant and intermediate modes.}

For non-resonant modes, we can estimate $\Delta \omega \sim 1/\tau _{0}$ and 
$\left\vert U_{\text{{\tiny AB}}}(t)\right\vert \sim 1/\hat{s}$ for any $%
t\gtrsim \tau _{0}$ --- see equations (\ref{NR-1}) and (\ref{A5U}). Hence,
decoherence of a moderate intensity $\tau _{d}>\tau _{0}$ leads to $%
W=\left\vert U_{\text{{\tiny AB}}}\right\vert ^{2}\sim 1/\hat{s}^{2}.$ The
characteristic frequency $\tilde{\omega}$ and time $\tilde{\tau}$ of
tunnelling associated with decoherence of non-resonant modes becomes 
\begin{equation}
\tilde{\omega}=\frac{1}{\tilde{\tau}}\approx \frac{1}{\tau _{d}\hat{s}^{2}}=%
\frac{\omega _{d}}{\hat{s}^{2}}  \label{w-n-d}
\end{equation}%
Decoherence promotes tunnelling carried by non-resonant modes and impedes
tunnelling conducted by resonant modes. While $\tilde{\omega}$ specified by (%
\ref{w-n-d})\ is generally smaller than $\tilde{\omega}$ given by (\ref%
{w-r-d})\ (assuming $\tau _{d}>\tau _{0}$), the non-resonant modes are
likely to be more numerous. The A-resonant modes are primarily responsible
for tunnelling from A to B and the B-resonant modes are primarily
responsible for tunnelling from B to A. The overall tunnelling rate is an
aggregate of the tunnelling rates produced by each mode and estimated by (%
\ref{w-n-d}). Note that the ratio of the number of A-resonant the number of
B-resonant modes is roughly proportional to \ $x_{\text{{\tiny A}}}/x_{\text{%
{\tiny B}}}$ for a given small energy interval; hence, in the equilibrium
(or near-equilibrium) conditions, where modes with close energies must have
similar amplitudes, the probability of finding a particle in a particular
section (e.g. A or B) is proportional to the volume of this section. Note
that Markov models (\ref{Markov2}) do not constrain the extent of tunnelling
by its unitary value $\varsigma ,$ but promote equidistribution between
modes.

The estimates for the intermediate modes are similar $\Delta \omega \sim
\left\vert \eta \right\vert /(\hat{s}\tau _{0})$ and $\left\vert U_{\text{%
{\tiny AB}}}(t)\right\vert \sim \left\vert \xi \right\vert \sim 1/\left\vert
\eta \right\vert $ for any $t\gtrsim \tau _{0}\hat{s}/\left\vert \eta
\right\vert ,$ where parameter $\eta =2\hat{s}\theta $ is moderately large $%
1\ll \left\vert \eta \right\vert =2\hat{s}\left\vert \theta \right\vert \ll 
\hat{s}$ and determines how far the mode is from the resonance and $%
\left\vert \eta \right\vert \sim \hat{s}$ corresponds to non-resonant modes.
The tunnelling rate depends on relative values of $\Delta \omega $ and $%
\omega _{d}$ 
\begin{equation}
\tilde{\omega}=\frac{1}{\tilde{\tau}}\approx \left\{ 
\begin{array}{cc}
\omega _{d}/\eta ^{2}, & \omega _{d}\leq \frac{\left\vert \eta \right\vert }{%
\tau _{0}\hat{s}} \\ 
\frac{1}{\tau _{0}^{2}\hat{s}^{2}\omega _{d}}, & \frac{1}{\tau _{0}}\geq
\omega _{d}\geq \frac{\left\vert \eta \right\vert }{\tau _{0}\hat{s}}%
\end{array}%
\right.  \label{w-d-i}
\end{equation}

\subsection{The effect of intensive decoherence.}

Finally as decoherence becomes more intensive and $\tau _{d}\lesssim \tau
_{0},$ the coherent solutions cannot be sustained within each section of the
box --- the model of standing and evolving waves gives way to quantum
particles represented by wave packets. The coherent solutions stretching
from one side of the section to another are meaningless if the
characteristic decoherence time is shorter than the time of reflection from
the walls. In these conditions we necessarily use the transmission $%
\left\vert q\right\vert ^{2}$ and reflection $\left\vert r\right\vert ^{2}$
probabilities associated with tunnelling, which are specified by (\ref%
{A1-ass}) for the case under consideration. There is no longer any
difference between the resonant and non-resonant modes. The probabilities of
location in section A and section B are governed by the following Markov
chain 
\begin{equation}
-\frac{dP_{\text{{\tiny A}}}}{dt}=\frac{dP_{\text{{\tiny B}}}}{dt}=\frac{%
u_{0}\left\vert q\right\vert ^{2}}{2}\left( \frac{P_{\text{{\tiny A}}}}{x_{%
\text{{\tiny A}}}}-\frac{P_{\text{{\tiny B}}}}{x_{\text{{\tiny B}}}}\right)
\end{equation}%
where the intensity of collisions with the barrier is evaluated to be
proportional to $u_{0}/(2x)$. Assuming $x_{\text{{\tiny A}}}\approx x_{\text{%
{\tiny B}}}\approx x_{0}$ With $\left\vert q\right\vert ^{2}$ given by (\ref%
{A1-ass}), the transmission frequency becomes 
\begin{equation}
\tilde{\omega}=\frac{1}{\tilde{\tau}}\approx \frac{u_{0}\left\vert
q\right\vert ^{2}}{x_{0}}=\frac{1}{\hat{s}^{2}\tau _{0}}  \label{w-t-d}
\end{equation}%
Note the consistency of (\ref{w-t-d}) with the previous estimates (\ref%
{w-r-d}) and (\ref{w-n-d}), which can be converted into (\ref{w-t-d}) by
substituting $\tau _{d}=\tau _{0}$. The model (\ref{w-t-d}) based on
tunnelling probabilities should be valid for a wide range of small
decoherence times $\tau _{d}\lesssim \tau _{0},$ perhaps as long as
decoherence does not interfere with the actual passage through \ the barrier.

\subsection{Intrinsic versus environmental decoherence\label{Sec6vs}}

\begin{figure}[h]
\begin{center}
\includegraphics[width=9cm,page=3,trim=0cm 0cm 0cm 2cm, clip ]{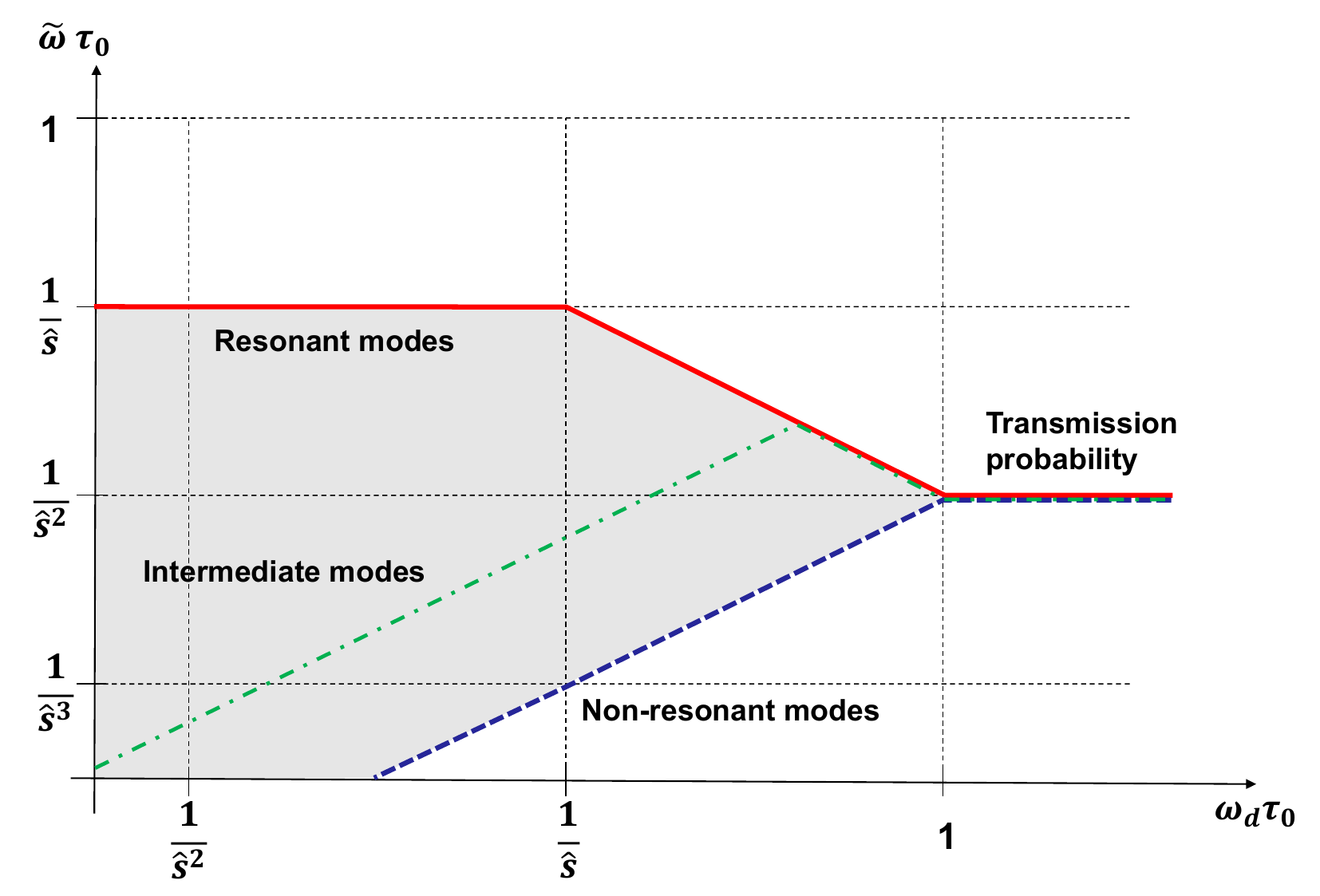}
\caption{The normalised rate of tunnelling depending on the normalised characteristic
decoherence rate for different modes. }
\label{fig3}
\end{center}
\end{figure}

The tunnelling frequencies are shown versus the decoherence frequency for
different modes in Figure \ref{fig3}. \ The tunnelling frequency of the
resonant modes decreases with increasing $\omega _{d}$ while the tunnelling
frequency of the non-resonant modes increases with increasing $\omega _{d}$
up until the both types of modes reach the common value specified by (\ref%
{w-t-d}). The figure also shows an intermediate mode that displays features
that are intermediate between the resonant and non-resonant modes. The
effect of intrinsic decoherence is complex but can be broadly characterised
by enhancing the extent of tunnelling and promoting equidistribution (and,
effectively, equilibration) of the particle locations between sections A and
B.

{\ Note that the effect of intrinsic (or effectively intrinsic) decoherence
on tunnelling considered here is generally different from the decoherence
effect produced by unitary interactions with a larger system or with the
environment. Unlike the former, the latter does not enhance the extent of
tunnelling. The effect of environmental interference does not become
significant until its energy of interactions becomes comparable with the
energy gap $E_{+}-E_{-}$. The effect of such significant decoherence is
conventionally described by the Zurek theory \cite{Zurek1982}, which
indicates that the rate of tunnelling increases significantly when the
energy of interactions exceeds the energy gap (see \ref{SecB}). If some
minor energy exchanges (much smaller than those required by thermalisation)
are allowed in addition to the classical interpretation of decoherence, then
the acceleration of tunnelling mentioned above would be supplemented by a
reduction of the extent of tunnelling, which may result in the effective
termination of tunnelling (see \ref{SecB} (\ref{SecBb})). One can see that
different types of decoherence affect tunnelling differently and, therefore,
can (at least in principle) be distinguished in experiments. }

\section{Discussion of the experiment \label{Sec7}}

{\ The Reichenbach conjecture suggests that all branch systems tend to
evolve forward in time towards equilibrated and thermalised conditions, even
if they are fully isolated from the rest of the universe. Obtaining
experimental confirmation or repudiation of this conjecture would be of
principal importance for our understanding of the universe. Thermalisation,
however, is the overall outcome of numerous microscopic processes, whose
fine mechanisms are concealed by the significance and magnitude of the
outcome. We, therefore, are interested in and focus on decoherence that, as
one would hope, can provide more information about the actual mechanisms of
time priming than thermalisation. We assume, by default, that the active
phase of the experiment $-t_{s}\leq t\leq +t_{s}$ is faster than the rate of
thermalisation $t_{s}\ll \tau _{t}.$ The thermalisation time $\tau _{t}$ can
be assessed during the passive phase of the experiment (e.g. by examining
the system after $t=+t_{s},$ when, as discussed in Section \ref{Sec4},
thermalisation is expected to screen the active phase of the experiment from
the final conditions, even if such final conditions are imposed on the
system by postselection). }

The present work analyses different regimes of interference between
decoherence and tunnelling, producing a range of behaviours illustrated in
Figure \ref{fig3} and in \ref{SecB}. The frequency of decoherence can be
estimated indirectly by measuring the tunnelling rates. Although some of
these experiments might be difficult to conduct, experimental studies of
decoherence \cite{ImryYoseph2002,Dec1Exp2008} and tunnelling \cite%
{Dattagupta2004,TunnZeno2014,QTG2018} that have some parallels with the
present analysis have been successfully carried out in the past. These
experiments, however, need to be modified to reduce influence of the
environment, avoid both thermalisation and near-zero temperatures, and
satisfy a number of conditions discussed below. As described in Section \ref%
{Sec3}, the suggested experiments involve trapping quantum particles in
section A, allowing them to tunnel to another section B and measuring the
tunnelling rates. This seems straight-forward but the devil is always in the
details.

In order to examine the effect of decoherence on tunnelling experimentally,
the characteristic times of tunnelling $\hat{\tau}\approx \hat{s}\tau _{0}=%
\hat{s}x_{0}/u_{0}$ and decoherence $\tau _{d}$ \ must be comparable. The
key point of the experiment is selecting experimental parameters so that the
transition between coherent and non-coherent regimes is observed. While the
characteristic time of tunnelling $\hat{\tau}$ can be changed in experiments
(although only within certain limits that are determined by the conditions
of the experiment), $\tau _{d}$ is expected to be very small for macroscopic
objects and very large (possibly infinite) for elementary particles. Hence,
the number of particles in the experiment needs to be selected so that the
expected decoherence rate for this system is not too large and not too small.

This experiment is concerned with the state of the quantum system when all
external interferences are (gradually) removed. As we increase the isolation
of the system by encircling it with perfect insulators, mirrors and shields,
screening the system from cosmic radiations and other forms of environmental
interferences, the intensity of environment-induced decoherence should also
reduce in proportion to the reduction of its cause. We might observe that at
some stage decoherence disappears or becomes too small and infrequent to be
detected --- in this case, we, as discussed above, need to increase the
scale of the experiment to bring the rate of decoherence into the measurable
range. If decoherence does not reappear even for sufficiently large,
macroscopic objects and decoherence can be reduced below any given level by
increasing isolation of a system, this would demonstrate the incorrectness
of the Reichenbach conjecture.

We assume, however, that the Reichenbach conjecture is correct and there is
a component of decoherence that cannot be eliminated by progressive
isolation of the system under any circumstances. We refer to such
ineliminable component as intrinsic (or effectively intrinsic). Measuring
the rate of tunnelling gives us information not only about the decoherence
rate but also about its nature. The effects of intrinsic and environmental
decoherences are similar in some respects but, as discussed in Section \ref%
{Sec6}(\ref{Sec6vs}), are different in others. One of the most interesting
outcomes of the experiments would be determining which of the two patterns
is followed by the ineliminable component of decoherence.

Any experiments that can bring some light into this matter and demonstrate
either existence of an (effectively) intrinsic component of decoherence or
its absence would be of the highest importance. The arrow of time is real
and so must be its time primer ---an underlying physical mechanism that
enacts the direction of time --- but, generally, it is difficult to say
whether this mechanism can be confidently detected under the current level
of technology.

The tunnelling experiments can be conducted with different particles:
photons, electrons, protons and, possibly, neutrons or even atomic nuclei
are the most likely candidates. The best choice of particles is not clear
--- while tunnelling is easier to achieve with lighter particles, photons
are expected to be decoherence-neutral \cite{Ent2017} and thus are less
likely to exhibit any intrinsic decoherence. Considering that the known
cases of CP violations, which have been detected in hadrons \cite%
{PDG2012,Barbar2016}, imply violation of the symmetry of time (assuming CPT
invariance) and that high-energy hadron collisions seem to lead to
thermodynamic behaviour in quark-gluon plasma \cite{Nature2007,MuB2011}, we
infer that these experiments point in the direction of protons and nuclei as
the most interesting particles for these experiments --- these particles are
most likely to possess properties associated with intrinsic decoherence,
presuming that such properties exist \cite{mixing2020}. (Note that
thermodynamic interferences may become apparent as ostensible CPT violations
in systems that are in fact CPT-preserving \cite{K-PhysA}.) The experiment
needs to be organised so that the tunnelling particles are baryons (or are
in contact with baryons although jointly isolated from the environment).
While cooling the surrounding to near-zero temperatures to control
environmental interferences seems like a good idea, cooling the system is
generally not desirable since this may dramatically reduce the magnitude of
intrinsic decoherence or completely freeze it.

If time-directional behaviour associated with decoherence can be detected in
the tunnelling of protons, it seems logical to conduct similar experiments
with antiprotons (assuming that the substantial practical difficulties
associated with such experiments can be overcome). Since conventional
thermodynamics can be extended from matter to antimatter in two possible
mutually exclusive ways: symmetric (i.e. CP-invariant) and antisymmetric
(i.e. CPT-invariant) \cite{KM-Entropy2014,SciRep2016,Ent2017}, decohering
behaviour of antiparticles is of particular interest. The CPT-invariant
version of thermodynamics expects antibaryons to predominantly recohere
while the CP-invariant version of thermodynamics insists that both baryons
and antibaryons must exhibit the same decohering behaviour.

\section{Conclusion \label{Sec8}}

The present work evaluates the effect of decoherence on the dynamic of
quantum tunnelling, carried out by resonant, intermediate and non-resonant
modes under generally non-equilibrium conditions that exist during the
active phase of the experiments. Decoherence tends to enhance tunnelling by
non-resonant modes and attenuate resonant tunnelling. The main conclusion of
the present analysis is that, under conditions considered here, the rate of
decoherence substantially affects the rate of tunnelling, and therefore can
be determined or estimated by measuring the rate of tunnelling. This seems
to be easier and less intrusive than direct testing of the coherent states.
The effects noted above become clear when the quantum barrier is high ($\hat{%
s}\gg 1$), the tunnelling transmission coefficient is low and the energy
eigenstates on both sides of the barrier are weakly coupled.

The problem of interference from the environment and measurements, which
inevitably cause decoherences and collapses, is especially pertinent to
examining decoherence. In simple terms, quantum measurements are bound to
cause the effects that they are intended to detect not create. Hence,
measuring the decoherence rates indirectly, through proxies is always
preferable. Examining tunnelling rates as proxies for decoherence rates and
using ancillary quantum systems to avoid direct interference seem very
useful in this context.

{\ This work shows that despite a significant degree of similarity, the
intrinsic (or effectively intrinsic) and unitary environmental mechanisms of
decoherence, affect the tunnelling rates differently and, therefore, can be,
at least in principle, experimentally distinguished from each other. In such
experiments, we need to minimise environmental interferences and avoid
strong interactions between different modes causing substantial energy
exchanges and thermalisation. }

The principal question that was formulated by Hans Reichenbach half a
century ago and still remains unanswered is whether thermodynamic
directionality of time would persist in fully isolated conditions. While
Reichenbach's conjecture (that it would) seems more probable, scientific
questions of this kind cannot be answered without experimental evidence. If
the arrow of time persists, there must be a dynamic mechanism (which we call
the time primer) that is responsible for this, and this mechanism should be
experimentally testable. This work suggests that these issues can be
examined in experiments involving quantum tunnelling.

\bigskip

\bigskip

{\ }

\section*{Declarations}

\subsection{Funding}

Not applicable

\subsection{Conflicts of interest / Competing interests}

The author states that there is no conflict of interest.

\subsection{Availability of data and material}

Not applicable

\subsection{Code availability (software application or custom code)}

Not applicable

\subsection{Authors' contributions}

Not applicable

\bigskip

\bigskip

\appendix{Tunnelling in a box and energy eigenstates \label{SecA}}

This Appendix presents equations for particle tunnelling in a rectangular
box and is subject to conditions imposed by the box boundaries --- the
problem is selected to allow for a complete and transparent analytical
evaluation. The results are used in the main body of the paper. Various
tunnelling solutions can be found in vast literature dedicated to this topic 
\cite{LL3,Tunn2003,Tunn2009}.

\subsection{Tunnelling through symmetric barriers\label{sec_A_tun}}

\begin{figure}[h]
\begin{center}
\includegraphics[width=6cm,page=4,trim=3cm 1.5cm 13cm 3cm, clip ]{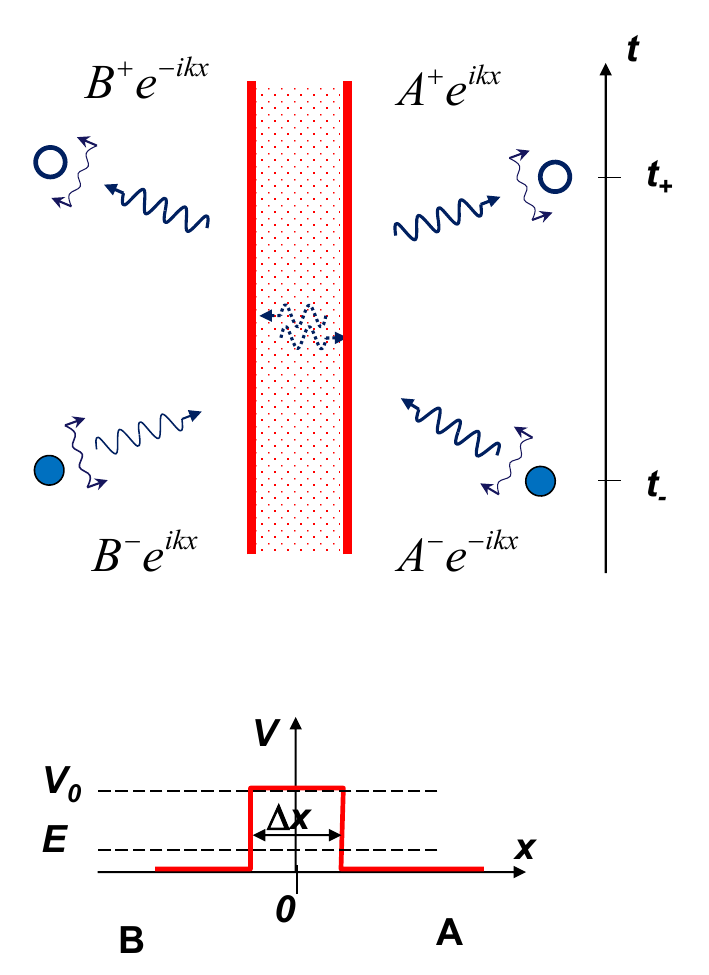}
\caption{Quantum tunnelling through a potential barrier: schematic of the incoming
and outgoing waves}
\label{fig4}
\end{center}
\end{figure}

The quantum outcomes of tunnelling can be expressed by the scattering matrix 
$\mathbb{S}$, which is a unitary matrix ($\mathbb{SS}^{\dag }\mathbb{=I}$)
that connects the amplitudes $A^{-}$ and\ $B^{-}$ of incoming waves $%
A^{-}e^{-i(\omega t+kx)}$ and\ $B^{-}e^{-i(\omega t-kx)}$ with the
amplitudes $A^{+}$ and\ $B^{+}$ of the outgoing waves $A^{+}e^{-i(\omega
t-kx)}$ and\ $B^{+}e^{-i(\omega t+kx)}$ (see Figure \ref{fig4}) so that:%
\begin{equation}
\left[ 
\begin{array}{c}
A^{+} \\ 
B^{+}%
\end{array}%
\right] =\mathbb{S}\left[ 
\begin{array}{c}
A^{-} \\ 
B^{-}%
\end{array}%
\right] ,\ \ \ \mathbb{S=}\left[ 
\begin{array}{cc}
\tilde{r} & \tilde{q} \\ 
-\tilde{q}^{\ast } & \tilde{r}^{\ast }%
\end{array}%
\right] =\left[ 
\begin{array}{cc}
r & q \\ 
q & r%
\end{array}%
\right]  \label{Scatter}
\end{equation}%
In the last expression for $\mathbb{S}$ in (\ref{Scatter}), the quantum
barrier is assumed to be symmetric, which corresponds to a symmetric matrix $%
\mathbb{S}$, which is invariant with respect swapping $A$ and $B$. The first
expression for $\mathbb{S}$ is general provided $\left\vert \tilde{q}%
\right\vert ^{2}+\left\vert \tilde{r}\right\vert ^{2}=1$. The reflection $r$
and transmission $q$ coefficients satisfy $\left\vert q\right\vert
^{2}+\left\vert r\right\vert ^{2}=1$ and $\left\vert r^{2}-q^{2}\right\vert
=1$ (implying that $\chi =q^{2}/r^{2}$ is real and $\chi \leq 0$) due to the
unitary of $\mathbb{S}$. Hence, $q=\pm ir(\left\vert r\right\vert
^{-2}-1)^{1/2}$ and $r=\mp iq(\left\vert q\right\vert ^{-2}-1)^{1/2}$.

The matrix $\mathbb{S}$ should not be confused with the commonly used
transfer matrix $\mathbb{M}$ that links the wave amplitudes on one side of
the barrier to the wave amplitudes on the other side. 
\begin{equation}
\left[ 
\begin{array}{c}
B^{-} \\ 
B^{+}%
\end{array}%
\right] =\mathbb{M}\left[ 
\begin{array}{c}
A^{+} \\ 
A^{-}%
\end{array}%
\right] ,\ \ \ \left[ 
\begin{array}{c}
A^{-} \\ 
A^{+}%
\end{array}%
\right] =\mathbb{M}\left[ 
\begin{array}{c}
B^{+} \\ 
B^{-}%
\end{array}%
\right] ,  \label{A1BA}
\end{equation}%
\ where 
\begin{equation}
\mathbb{M=}\frac{1}{q}\left[ 
\begin{array}{cc}
1 & -r \\ 
r & q^{2}-r^{2}%
\end{array}%
\right]  \label{A1M}
\end{equation}%
and $q^{2}-r^{2}=-r^{2}/\left\vert r^{2}\right\vert =q^{2}/\left\vert
q^{2}\right\vert $.

The values of\ $r$ and $q$ can be easily evaluated for a rectangular barrier
of height $V_{0}$ and width $\Delta x$ \cite{LL3,mixing2020}. Assuming that $%
V_{0}\rightarrow \infty $ and $\Delta x\rightarrow 0$ so that $s\
=V_{0}\Delta x\ \sim \func{const}$ and $V(x)\rightarrow s\delta (x),$ we
obtain 
\begin{equation}
q=\frac{1}{1+i\hat{s}},\ \ r=\frac{-i\hat{s}}{1+i\hat{s}},\   \label{A1qr}
\end{equation}%
where 
\begin{equation}
\hat{s}=\frac{\tilde{s}}{k}=\frac{\kappa ^{2}\Delta x}{2k}=\frac{m}{k\hbar
^{2}}s=\frac{V_{0}}{\hbar }\frac{\Delta x}{u_{0}},\ \ \ \kappa ^{2}=\frac{2m%
}{\hbar ^{2}}V_{0},\ \ \ s=V_{0}\Delta x,\ \ u_{0}=\frac{k\hbar }{m}
\label{A1par}
\end{equation}%
If $A^{+}=\left( A^{-}\right) ^{\ast }=A,$ then (\ref{A1BA})-(\ref{A1qr})
yield $B^{+}=\left( B^{-}\right) ^{\ast }=B=A^{\ast }-i\hat{s}(A+A^{\ast })$
and 
\begin{equation}
A+A^{\ast }=B+B^{\ast },\ \ \ \text{ }B-B^{\ast }+A-A^{\ast }+2i\hat{s}%
(A+A^{\ast })=0  \label{A1jmp}
\end{equation}%
With $\left\vert q\right\vert ^{2}$ ranging from 1 to 0 and $\left\vert
r\right\vert ^{2}$ ranging from 0 to 1 as $\hat{s}$ increases from 0 to $%
\infty $, the barrier shaped as the delta function is a basic representation
for many other barriers. Generally, $r$ and $q$ can be jointly multiplied by
any arbitrary phase $e^{i\vartheta _{1}}$ and preserve unitarity of $\mathbb{%
S}$ (if the barrier is non-symmetric, then $\mathbb{S}$ involves another
arbitrary angle $\vartheta _{2}$) but, if the phase shifts are not of major
concern, the delta function tends to provide a good model for interactions
of a wave function of given $k$ with the barriers.

If $\hat{s}\rightarrow \infty $, the transmission $\left\vert q\right\vert
^{2}$ and reflection $\left\vert r\right\vert ^{2}$ probabilities are given
by 
\begin{equation}
\left\vert q\right\vert ^{2}=\frac{1}{\hat{s}^{2}},\ \left\vert r\right\vert
^{2}=1-\frac{1}{\hat{s}^{2}}  \label{A1-ass}
\end{equation}%
These equations are special cases of more general expressions for the
transmission and reflection probabilities obtained by Igor Vladimirov (2008,
unpublished).

\subsection{Energy eigenstates}

The eigenstates of the Schr\"{o}dinger equation (\ref{eig1x}) 
\begin{equation}
\mathbb{\tilde{H}}\tilde{\Psi}_{j}=-\frac{\hbar ^{2}}{2m}\frac{\partial ^{2}%
\tilde{\Psi}_{j}}{\partial x^{2}}+V(x)\tilde{\Psi}_{j}=\tilde{E}_{j}\tilde{%
\Psi}_{j}  \label{A2eq}
\end{equation}%
are to be determined within the interval $-x_{\text{{\tiny B}}}\leq x\leq x_{%
\text{{\tiny A}}}$ with homogeneous boundary conditions 
\begin{equation}
\tilde{\Psi}_{j}=0\ \ \text{at\ \ }x=x_{\text{{\tiny A}}}\ \ \text{and\ \ }%
\tilde{\Psi}_{j}=0\ \ \text{at\ \ }x=-x_{\text{{\tiny B}}}  \label{A2bc}
\end{equation}%
and singular potential $V(x)=s\delta (x)$. The parameter $s$ is assumed to
be sufficiently large so that the probabilities of tunnelling through the
barrier are low.

\subsection{Note on singular potentials}

Consider a rectangular barrier $V=V_{0}$ at $-\Delta x/2\leq x\leq +\Delta
x/2$ and $V=0$ elsewhere. The limit $V_{0}=V_{n}^{\circ }\rightarrow \infty
, $ $\Delta x=\Delta x_{n}\rightarrow 0$ as $n=1,2,...$ so that $%
V_{n}^{\circ }\Delta x_{n}=s$ corresponds to introducing singularity $%
V(x)=V_{n}(x)\rightarrow s\delta (x)$ into the model. The presence of the
delta function $\delta (x)$ in the potential does not affect validity of the
Hilbert--Schmidt theorem. With the use of the Green function $\mathbb{\tilde{%
H}}G(x,x_{0})=\delta (x-x_{0})$ and $G=0\ $at $x=x_{\text{{\tiny A}}}\ $and\ 
$x=x_{\text{{\tiny B}}}$,\ the eigenstate problem $\mathbb{\tilde{H}}\tilde{%
\Psi}_{j}=\tilde{E}_{j}\tilde{\Psi}_{j}$ is conventionally converted into a
Fredholm integral equation 
\begin{equation}
\tilde{\Psi}_{j}(x)=\tilde{E}_{j}\mathbb{G}\tilde{\Psi}_{j}=\tilde{E}%
_{j}\dint\limits_{-x_{\text{{\tiny B}}}}^{x_{\text{{\tiny A}}}}G(x,x_{0})%
\tilde{\Psi}_{j}(x_{0})dx_{0}
\end{equation}%
where the integral operator $\mathbb{G=\tilde{H}}^{-1}$ is compact and
Hermitian in compliance with the conditions of the Hilbert--Schmidt theorem.
In three-dimensional case, the Green function defined by $\mathbb{H}G(%
\mathbf{r},\mathbf{r}_{0})=\delta (\mathbf{r}-\mathbf{r}_{0})$ and $G=0\ $at 
$\mathbf{r}\in \partial $AB can be used to convert the eigenstate problem $%
\mathbb{H}\Psi _{j}=E_{j}\Psi _{j}$ into integral equation. Since the
sequence $\mathbb{G}_{n}$ of integral operators $\mathbb{G}_{1},\mathbb{G}%
_{2},...$ corresponding to $V_{0}=V_{1}^{\circ },V_{2}^{\circ },...$
converge $\mathbb{G}_{n}\rightarrow \mathbb{G}_{\delta }$ by the operator
norm when $n\rightarrow \infty $ and $V_{n}(x)\rightarrow s\delta (x)$, the\
theorem by Kolmogorov and Fomin \cite{KolmogorovFomin} (Theorem 1, Sec. 2,
Chpt. 6, Part IV ) ensures that the limiting integral operator $\mathbb{G}%
_{\delta }$ is compact and, obviously, Hermitian. Hence, solution (\ref{1sol}%
) must be universally valid even for singular potentials $V(x)=s\delta (x)$.
The system of energy eigenstates is complete in Hilbert space and covers all
possible evolutions of the Schr\"{o}dinger equation.

\subsection{Eigenfunctions for a delta-function barrier}

Since Hamiltonian $\mathbb{\tilde{H}}$ is time-symmetric, the energy
eigenstates $\tilde{\Psi}_{j}$ can be treated as real without loss of
generality. Assuming $V(x)=s\delta (x),$ the solution of (\ref{A2eq}) with
boundary conditions (\ref{A2bc}) is given by 
\begin{equation}
\tilde{\Psi}_{j}=\left\{ 
\begin{array}{c}
A_{j}\sin (k_{j}x+\alpha _{j}),\ \ \ \ \alpha _{j}=-k_{j}x_{\text{{\tiny A}}}%
\text{ \ \ in section A} \\ 
B_{j}\sin (k_{j}x+\beta _{j}),\ \ \ \ \beta _{j}=+k_{j}x_{\text{{\tiny B}}}%
\text{ \ \ in section B}%
\end{array}%
\right.  \label{3PSI}
\end{equation}%
The amplitudes $A_{j}$ and $B_{j},$ which are assumed real, are constrained
by (\ref{A1jmp}) (i.e. $2A=-iA_{j}e^{i\alpha _{j}}$ and $2B=-iB_{j}e^{i\beta
_{j}}$), that is by continuity of the functions $\ x=0$ and jumps of the
derivatives induced by $V=s\delta (x)$ 
\begin{equation}
B_{j}\sin (\beta _{j})=A_{j}\sin (\alpha _{j}),\ \ \ B_{j}\cos (\beta
_{j})=A_{j}\cos (\alpha _{j})-2\frac{\tilde{s}}{k_{j}}A_{j}\sin (\alpha _{j})
\label{3BA}
\end{equation}%
Dividing the second equation by the first equation and substituting $\alpha
_{j}$ and $\beta _{j}$ from (\ref{3PSI}) yields the dispersion equation 
\begin{equation}
\cot (k_{j}x_{\text{{\tiny B}}})+\cot (k_{j}x_{\text{{\tiny A}}})+2\frac{%
\tilde{s}}{k_{j}}=0  \label{3disp}
\end{equation}%
that determines energy eigenvalues 
\begin{equation}
\tilde{E}_{j}=\frac{k_{j}^{2}\hbar ^{2}}{2m}  \label{3Ej}
\end{equation}%
in terms of $\tilde{s}=ms/\hbar ^{2}$. The amplitude ratio is then given by 
\begin{equation}
\frac{B_{j}}{A_{j}}=-\frac{\sin (k_{j}x_{\text{{\tiny A}}})}{\sin (k_{j}x_{%
\text{{\tiny B}}})}  \label{3rat}
\end{equation}%
In the rest of the analysis we assume that $\hat{s}=\tilde{s}/k$ is large
for typical values of $k$ to simplify the equations and obtain conditions
that are of interest for our consideration.

\subsection{The resonance case\label{sec_res}}

In this case $x_{\text{{\tiny B}}}=x_{\text{{\tiny A}}}=x_{0}$ and all modes
are resonant. Two family of solutions are distinguished: first,
antisymmetric $\tilde{\Psi}_{j}(-x)=-\tilde{\Psi}_{j}(x),$ smooth at $x=0$
with $k_{j}$ specified by $\sin (k_{j}^{(\text{a})}x_{0})=0;$ and, second,
symmetric $\tilde{\Psi}_{j}(-x)=\tilde{\Psi}_{j}(x),$ V-shaped at $x=0$ with 
$k_{j}$ evaluated from $\cot (k_{j}^{(\text{s})}x_{0})=-\tilde{s}/k_{j}^{(%
\text{s})}$. Note that $\left\vert B_{j}\right\vert =\left\vert
A_{j}\right\vert $ for both of the families. Assuming that $\hat{s}=\tilde{s}%
/k_{j}$ is large, we expand $\cot (\pi j+\alpha )=1/\alpha +...$ and obtain 
\begin{equation}
k_{j}^{(\text{a})}=\frac{\pi j}{x_{0}},\ \ \ k_{j}^{(\text{s})}\approx \frac{%
\pi j}{x_{0}}\left( 1-\frac{1}{x_{0}\tilde{s}}\right) ,\ \ \ \left( \frac{%
B_{j}}{A_{j}}\right) ^{(\text{a})}=1,\ \ \ \left( \frac{B_{j}}{A_{j}}\right)
^{(\text{s})}=-1  \label{A2res}
\end{equation}%
where$\ j=1,$ $2,$ $3,$ $...$ for both the symmetric (s) and antisymmetric
(a) modes.

Existence of symmetric and antisymmetric modes is a general property of
quantum equations with any symmetric potential $V(x)=V(-x)$ (implying that $%
x_{\text{{\tiny B}}}=x_{\text{{\tiny A}}}$). Indeed, let $\tilde{\Psi}_{j}$
be a solution of (\ref{A2eq}) and $\mathbb{P}$ be the parity operator that
transforms $x\rightarrow -x$. Without loss of generality we can assume that $%
\tilde{\Psi}_{j}$ is real. The parity transformation preserves (\ref{A2eq})
for symmetric potentials $V(x)$ and $[\mathbb{P},\mathbb{\tilde{H}}]=0.$
Hence, $\mathbb{P}\tilde{\Psi}_{j}$ is also solution of (\ref{A2eq}) and,
provided the eigenvalue $\tilde{E}_{j}$ is not degenerate, we can always
chose $c$ so that $c\mathbb{P}\tilde{\Psi}_{j}=\tilde{\Psi}_{j}$ coincides
with the original solution, where $c$ is an unknown constant satisfying $%
\left\vert c\right\vert =1$ to preserve normalisation. By applying the
operator $c\mathbb{P}$ twice we obtain $x\rightarrow x$ and $c\mathbb{P}c%
\mathbb{P}\tilde{\Psi}_{j}=c^{2}\tilde{\Psi}_{j}=\tilde{\Psi}_{j}$. Hence,
either $c=+1,$ which corresponds to a symmetric mode, or $c=-1$, which
corresponds to an antisymmetric mode. Under the limit of a high,
impenetrable barrier (i.e. $\hat{s}\rightarrow \infty $ in our terms) the
wave functions in sections A and B interact less and less and, therefore,
the symmetric and antisymmetric modes become very similar and merge $\tilde{E%
}_{j}^{(\text{a})}-\tilde{E}_{j}^{(\text{s})}\rightarrow 0$.

\subsection{Non-resonant modes \label{sec_non_res}}

If $x_{\text{{\tiny B}}}\neq x_{\text{{\tiny A}}}$, at least some and,
typically, most modes are non-resonant. Assuming that $\hat{s}=\tilde{s}/k$
is large, we identify two family of solutions among the non-resonant modes:
A-resonant where $\cot (k_{j}x_{\text{{\tiny A}}})\approx -2\tilde{s}/k_{j}$
and B-resonant where $\cot (k_{j}x_{\text{{\tiny B}}})\approx -2\tilde{s}%
/k_{j}$. For these modes, one can easily obtain from (\ref{3disp}) and (\ref%
{3rat}) the following expansions 
\begin{equation}
k_{j}^{(\text{{\tiny A}})}\approx \frac{\pi j}{x_{\text{{\tiny A}}}}\left( 1-%
\frac{1}{2x_{\text{{\tiny A}}}\tilde{s}}\right) ,\text{ }\ \left( \frac{B_{j}%
}{A_{j}}\right) ^{(\text{{\tiny A}})}\approx \sigma _{j}\frac{\pi j/(2x_{%
\text{{\tiny A}}}\tilde{s})}{\sin (\pi jx_{\text{{\tiny B}}}/x_{\text{{\tiny %
A}}})}\sim \frac{1}{\hat{s}}\ll 1  \label{A2Ar}
\end{equation}%
\begin{equation}
k_{j}^{(\text{{\tiny B}})}\approx \frac{\pi j}{x_{\text{{\tiny B}}}}\left( 1-%
\frac{1}{2x_{\text{{\tiny B}}}\tilde{s}}\right) ,\text{ }\ \left( \frac{B_{j}%
}{A_{j}}\right) ^{(\text{{\tiny B}})}\approx \sigma _{j}\frac{\sin (\pi jx_{%
\text{{\tiny A}}}/x_{\text{{\tiny B}}})}{\pi j/(2x_{\text{{\tiny B}}}\tilde{s%
})}\sim \hat{s}\gg 1  \label{A2Br}
\end{equation}%
where $\ j=1,$ $2,$ $3,$ $...$ and $\sigma _{j}=\cos (\pi j)=(-1)^{j}$
alternates the signs. These expressions are valid unless a mode is (or is
close to) A-resonant and B-resonant at the same time --- these resonant,
near-resonant or intermediate modes require a more careful examination and
are considered below.

\subsection{Resonant, near-resonant and intermediate modes\label%
{sec_near_res}}

Although $x_{\text{{\tiny B}}}\neq x_{\text{{\tiny A}}}$ some of the modes
can still be exactly resonant or close to resonant conditions simultaneously
in both sections A and B: \ 
\begin{equation}
k_{0}x_{\text{{\tiny A}}}=\pi j_{\text{{\tiny A}}}-\theta \text{ \ \ and \ \ 
}k_{0}x_{\text{{\tiny B}}}=\pi j_{\text{{\tiny B}}}+\theta  \label{A2rm-jAjB}
\end{equation}%
for some real $k_{0},$ integer $j_{\text{{\tiny A}}}$ and integer $j_{\text{%
{\tiny B}}},$ where $\left\vert \theta \right\vert \sim 1/\hat{s}\ll 1$ is a
phase shift indicating small deviations from the resonance. The condition $%
\theta =0$ corresponds to exact resonance. In the rest of the Appendix the
subscript "$j$" is omitted implying that wave vectors, energies and
amplitudes considered here are related to a selected mode with some integer $%
j_{\text{{\tiny A}}}$ and $j_{\text{{\tiny B}}}$ in (\ref{A2rm-jAjB}). Let $%
k=k_{0}+\Delta k$ where $\Delta k\sim 1/\hat{s}$ is small, then at the
leading order 
\begin{equation}
\frac{1}{x_{\text{{\tiny B}}}\Delta k+\theta }+\frac{1}{x_{\text{{\tiny A}}%
}\Delta k-\theta }+2\hat{s}=0  \label{A2rm-eq1}
\end{equation}%
\begin{equation}
\frac{B}{A}=-\sigma \frac{x_{\text{{\tiny A}}}\Delta k-\theta }{x_{\text{%
{\tiny B}}}\Delta k+\theta },\ \ \sigma =\frac{\cos (\pi j_{\text{{\tiny A}}%
})}{\cos (\pi j_{\text{{\tiny B}}})}=\pm 1  \label{A2rm-eq2}
\end{equation}%
Equations (\ref{A2rm-eq1})-(\ref{A2rm-eq2}) can be solved to yield:%
\begin{equation}
\Delta k_{\mp }=\frac{1}{4\hat{s}}\frac{\left( \eta -1\right) x_{\text{%
{\tiny B}}}-\left( \eta +1\right) x_{\text{{\tiny A}}}\mp D^{1/2}}{x_{\text{%
{\tiny A}}}x_{\text{{\tiny B}}}},\ \ \ E_{\mp }=\frac{k_{0}\hbar ^{2}}{2m}%
(k_{0}+2\Delta k_{\mp })  \label{A2rm-dk}
\end{equation}%
\begin{equation}
\left( \frac{B}{A}\right) _{\mp }=\sigma \frac{x_{\text{{\tiny A}}}}{x_{%
\text{{\tiny B}}}}F_{\mp }\left( \eta ,\frac{x_{\text{{\tiny B}}}}{x_{\text{%
{\tiny A}}}}\right) ,\ \ \ F_{\mp }=\frac{(q_{+})\pm D^{1/2}}{(q_{-})\mp
D^{1/2}},\ \ \ \ \eta =2\hat{s}\theta  \label{A2rm-BA}
\end{equation}%
where%
\begin{equation}
q_{\pm }=\left( \eta \pm 1\right) (x_{\text{{\tiny A}}}+x_{\text{{\tiny B}}%
}),\ \ \ D=(x_{\text{{\tiny A}}}+x_{\text{{\tiny B}}})\left( \left( \eta
-1\right) ^{2}x_{\text{{\tiny B}}}+\left( \eta +1\right) ^{2}x_{\text{{\tiny %
A}}}\right)  \label{A2rm-D}
\end{equation}%
Note the equality 
\begin{equation}
\left( \frac{B}{A}\right) _{-}\left( \frac{B}{A}\right) _{+}=-\frac{x_{\text{%
{\tiny A}}}}{x_{\text{{\tiny B}}}}  \label{A2rm-rat}
\end{equation}%
which implies that when one branch of the solution becomes large, the other
inevitably becomes small and vice versa. The superscript indices "$+$" and "$%
-$" are used to denote values that correspond to the "plus" and "minus"
solutions of (\ref{A2rm-eq1}). When the sections of the box are of similar
sizes $x_{\text{{\tiny B}}}\approx x_{\text{{\tiny A}}}\approx x_{0}$
(although not necessarily exactly identical $x_{\text{{\tiny B}}}\neq x_{%
\text{{\tiny A}}}$), equations (\ref{A2rm-dk}) and (\ref{A2rm-BA}) can be
simplified 
\begin{equation}
\Delta k_{\mp }\approx \frac{1}{2\hat{s}}\frac{-1\mp \sqrt{(\eta ^{2}+1)}}{%
x_{0}},\ \ \left( \frac{B}{A}\right) _{\mp }\approx \sigma \frac{\left( \eta
+1\right) \pm \sqrt{(\eta ^{2}+1)}}{\left( \eta -1\right) \mp \sqrt{(\eta
^{2}+1)}}
\end{equation}

\subsection{Asymptotes for the resonant and intermediate modes}

When using parameter $\eta $, we distinguish resonant $\eta \rightarrow 0,$
near-resonant $\left\vert \eta \right\vert \sim 1$, intermediate $1\ll
\left\vert \eta \right\vert \ll \hat{s}$ and non-resonant $\left\vert \eta
\right\vert \sim \hat{s}\gg 1$ modes. For equations (\ref{A2rm-dk}) and (\ref%
{A2rm-BA}), the resonance limit of $\eta =2\hat{s}\theta \rightarrow 0$ is
given by 
\begin{equation}
\Delta k_{-}=-\frac{1}{2\hat{s}}\frac{x_{\text{{\tiny A}}}+x_{\text{{\tiny B}%
}}}{\ x_{\text{{\tiny A}}}x_{\text{{\tiny B}}}}-\frac{x_{\text{{\tiny A}}%
}-x_{\text{{\tiny B}}}}{x_{\text{{\tiny A}}}x_{\text{{\tiny B}}}}\eta +...\
,\ \ \ \ \Delta k_{+}=\frac{1}{2\hat{s}}\frac{\eta ^{2}}{x_{\text{{\tiny A}}%
}+x_{\text{{\tiny B}}}}+...  \label{A4dk}
\end{equation}%
\begin{equation}
\left( \frac{B}{A}\right) _{-}=-\sigma \frac{x_{\text{{\tiny A}}}}{x_{\text{%
{\tiny B}}}}\left( 1+\eta \right) +...\ ,\ \ \ \left( \frac{B}{A}\right)
_{+}=\sigma \left( 1-\eta \right) +...  \label{A4BA}
\end{equation}%
Comparison with the resonance case of subsection (\ref{sec_res}) indicates
that, at $\eta =0$ and $x_{\text{{\tiny B}}}=x_{\text{{\tiny A}}},$ the
"minus" solution represents the symmetric mode and the "plus" solution
represents the antisymmetric mode.

The asymptotic representation of equations (\ref{A2rm-dk}) and (\ref{A2rm-BA}%
) for intermediate modes is evaluated at the non-resonant limit $\eta
\rightarrow +\infty $ yielding%
\begin{equation}
\Delta k_{-}=-\frac{1}{2\hat{s}}\frac{\eta +1}{x_{\text{{\tiny B}}}}+...\ ,\
\ \ \ \Delta k_{+}=\frac{1}{2\hat{s}}\frac{\eta -1}{x_{\text{{\tiny A}}}}+...
\label{A2dk_inf}
\end{equation}%
\begin{equation}
\left( \frac{B}{A}\right) _{-}=-\sigma \frac{x_{\text{{\tiny A}}}+x_{\text{%
{\tiny B}}}}{x_{\text{{\tiny B}}}}\eta +...\ ,\ \ \ \left( \frac{B}{A}%
\right) _{+}=\sigma \frac{x_{\text{{\tiny A}}}}{x_{\text{{\tiny A}}}+x_{%
\text{{\tiny B}}}}\frac{1}{\eta }+...  \label{A2BA_inf}
\end{equation}%
The "minus" branch matches the B-resonant solution in (\ref{A2Br}) and the
"plus" branch matches the A-resonant solution in (\ref{A2Ar}). For example,
substituting $\pi j_{\text{{\tiny B}}}=\pi j_{\text{{\tiny A}}}x_{\text{%
{\tiny B}}}/x_{\text{{\tiny A}}}-\left( 1+x_{\text{{\tiny B}}}/x_{\text{%
{\tiny A}}}\right) \theta $ obtained from (\ref{A2rm-jAjB}) into $\sin (\pi
jx_{\text{{\tiny A}}}/x_{\text{{\tiny B}}})$ in (\ref{A2Br}) (while putting $%
j=j_{\text{{\tiny B}}}$\ and expanding $\sin (...)$ to the leading order)\
results in the first equation in (\ref{A2BA_inf}). Similarly, substituting
the equivalent expression $\pi j_{\text{{\tiny A}}}=\pi j_{\text{{\tiny B}}%
}x_{\text{{\tiny A}}}/x_{\text{{\tiny B}}}+\left( 1+x_{\text{{\tiny A}}}/x_{%
\text{{\tiny B}}}\right) \theta $ into expansion of $\sin (\pi jx_{\text{%
{\tiny B}}}/x_{\text{{\tiny A}}})$ in (\ref{A2Ar}) (while putting $j=j_{%
\text{{\tiny A}}}$\ this time)\ results in the second equation in (\ref%
{A2BA_inf}). \ These asymptotes, however, are swapped under the limit $\eta
\rightarrow -\infty $ that yields the following expressions: 
\begin{equation}
\Delta k_{-}=\frac{1}{2\hat{s}}\frac{\eta -1}{x_{\text{{\tiny A}}}}+...\ ,\
\ \ \ \Delta k_{+}=-\frac{1}{2\hat{s}}\frac{\eta +1}{x_{\text{{\tiny B}}}}%
+...
\end{equation}%
\begin{equation}
\left( \frac{B}{A}\right) _{-}=\sigma \frac{x_{\text{{\tiny A}}}}{x_{\text{%
{\tiny A}}}+x_{\text{{\tiny B}}}}\frac{1}{\eta }+...\ ,\ \ \ \left( \frac{B}{%
A}\right) _{+}=-\sigma \frac{x_{\text{{\tiny A}}}+x_{\text{{\tiny B}}}}{x_{%
\text{{\tiny B}}}}\eta +...
\end{equation}

{\ \appendix{Tunnelling influenced by environmental interferences \label%
{SecB}} }

The system under consideration is placed in a contact with a larger system
or the environment, which has the following energy eigenstates $\left\vert
l\right\rangle =\left\vert E_{l}\right\rangle ,$ where $l=1,2,...,N_{e}$ and 
$N_{e}$ is extremely large. The joint Hamiltonian of the system and
environment is given by the usual expression 
\begin{equation}
\mathbb{H=H}_{s}\mathbb{\otimes I}_{e}+\mathbb{I}_{s}\mathbb{\otimes H}_{e}+%
\mathbb{H}_{\text{int}}  \label{HsHint}
\end{equation}%
where the subscripts "$s$" indicates the system, "$e$" indicates the
environment (or a larger system) and $\mathbb{H}_{\text{int}}$ specifies
interactions between the system and the environment and acts in the system $%
\mathbb{\otimes }$ environment product space $\left\vert s\right\rangle
\left\vert l\right\rangle =\left\vert s\right\rangle \mathbb{\otimes }%
\left\vert l\right\rangle $. We consider weak interactions of the
eigenstates of the system Hamiltonian $\mathbb{H}_{s}=\mathbb{H}_{0}+\mathbb{%
H}_{1}$, which is specified previously in (\ref{H0H1}), with a selected
environmental energy eigenstate $\left\vert l\right\rangle$. Obtaining the
overall solution of the problem $\left\vert \psi _{s\mathbb{\otimes }%
e}\right\rangle $ is followed by tracing out the degrees of freedom
associated with the environment to determine the effective density matrix of
the system: $\mathbf{\rho }_{s}=\func{tr}_{e}(\left\vert \psi _{s\mathbb{%
\otimes }e}\right\rangle \left\langle \psi _{s\mathbb{\otimes }e}\right\vert
).$ In our analysis, we omit the environmental energy exponents $\exp \left(
-iE_{l}t^{\circ }/\hbar \right) $ since they do not affect the trace. As in
other theories of environmental decoherence, we necessarily use antecedent
causality in this analysis.

\subsection{Environmental decoherence without energy exchange \label{SecBa}}

According to Zurek's theory \cite{Zurek1982}, decoherence occurs due to
environmental interferences without any energy exchange between different
energy modes. Hence, the interaction Hamiltonian takes a diagonal form when
energy eigenstates are used so that 
\begin{equation}
\left\langle +\right\vert \left\langle l\right\vert \mathbb{H}_{\text{int}%
}\left\vert +\right\rangle \left\vert l\right\rangle =E_{+l},\ \ \ \
\left\langle -\right\vert \left\langle l\right\vert \mathbb{H}_{\text{int}%
}\left\vert -\right\rangle \left\vert l\right\rangle =E_{-l}
\end{equation}%
and all other components are zeros, for example, $\left\langle -\right\vert
\left\langle l\right\vert \mathbb{H}_{\text{int}}\left\vert +\right\rangle
\left\vert l\right\rangle =0$ and $\left\langle +\right\vert \left\langle
l\right\vert \mathbb{H}_{\text{int}}\left\vert +\right\rangle \left\vert
l^{\prime }\right\rangle =0$ when $l\neq l^{\prime }$. This form of the
interaction Hamiltonian $\mathbb{H}_{\text{int}}$ in (\ref{HsHint}) results
in adjustments of the natural frequencies of the system (i.e. $\omega _{+}$
and $\omega _{-}$). The solution of this problem is obvious and, by analogy
with (\ref{ABt}), is given by 
\begin{equation}
\left[ 
\begin{array}{c}
\tilde{A} \\ 
\tilde{B}%
\end{array}%
\right] _{l}=\frac{e^{-i(\omega _{0}+\omega _{0l})t^{\circ }}}{1+\xi ^{2}}%
\left[ 
\begin{array}{c}
\exp \left( -i\frac{\Delta \omega +\Delta \omega _{_{l}}}{2}t^{\circ
}\right) +\xi ^{2}\exp \left( +i\frac{\Delta \omega +\Delta \omega _{_{l}}}{2%
}t^{\circ }\right) \\ 
-2i\xi \sin \left( \frac{\Delta \omega +\Delta \omega _{_{l}}}{2}t^{\circ
}\right)%
\end{array}%
\right]
\end{equation}%
where $\Delta \omega _{_{l}}=(E_{+l}-E_{-l})/\hbar $ and $\omega
_{0l}=(E_{+l}+E_{-l})/(2\hbar ),$ while the other quantities $\omega _{0},$ $%
\Delta \omega $ and $\xi $ are the same as defined in (\ref{A5eq1}), (\ref%
{A5eq2}) and (\ref{A5AB}). One can easily see that the effect of environment
is negligible as long as $\Delta \omega _{_{l}}\ll \Delta \omega $. If,
however, $\Delta \omega _{_{l}}\gg \Delta \omega ,$ then the environment
would cause a rapid loss of the coherence between the "plus" $\left\vert
+\right\rangle \left\vert l\right\rangle $ and "minus" $\left\vert
-\right\rangle \left\vert l\right\rangle $ modes, resulting in the
corresponding acceleration of tunnelling without any changes in the extent
of tunnelling determined by $\varsigma =\left\vert \xi \right\vert /(1+\xi
^{2}) $.

\subsection{Environmental decoherence with minimal energy exchanges\label%
{SecBb}}

The analysis of the influence of decoherence on tunnelling considered in
Section \ref{Sec6} uses the partition states $\left\vert \text{A}%
\right\rangle $ and $\left\vert \text{B}\right\rangle $ as the decoherence
basis. We now apply a similar assumption, implying that the environmental
interferences affect sections A and B autonomously. As demonstrated below,
it is sufficient to assume that the environment interferes only with the
state $\left\vert \text{A}\right\rangle $ but not with the state $\left\vert 
\text{B}\right\rangle$. As in Section \ref{Sec6}, the energy exchanges due
to these interferences are deemed to be small (i.e. weaker than those that
can cause thermalisation during the active stage of the experiment). The
Hamiltonian takes the following form 
\begin{equation}
\left\langle \text{A}\right\vert \left\langle l\right\vert \mathbb{H}_{\text{%
int}}\left\vert \text{A}\right\rangle \left\vert l\right\rangle =E_{\text{A}%
l}  \label{AppB_H}
\end{equation}%
while the other elements of the interaction Hamiltonian are zeros. Note that
simultaneous adjustment of both energies $\left\langle \text{A}\right\vert
\left\langle l\right\vert \mathbb{H}_{\text{int}}\left\vert \text{A}%
\right\rangle \left\vert l\right\rangle =E_{0l}$ and $\left\langle \text{B}%
\right\vert \left\langle l\right\vert \mathbb{H}_{\text{int}}\left\vert 
\text{B}\right\rangle \left\vert l\right\rangle =E_{0l}$ corresponds to $%
\left\langle +\right\vert \left\langle l\right\vert \mathbb{H}_{\text{int}%
}\left\vert +\right\rangle \left\vert l\right\rangle =E_{0l}$ and $%
\left\langle -\right\vert \left\langle l\right\vert \mathbb{H}_{\text{int}%
}\left\vert -\right\rangle \left\vert l\right\rangle =E_{0l}$ and results in
a mere adjustment of the principal frequency $\omega _{0}$. Any exclusive
interference of the environment with section B $\left\langle \text{B}%
\right\vert \left\langle l\right\vert \mathbb{H}_{\text{int}}\left\vert 
\text{B}\right\rangle \left\vert l\right\rangle =E_{\text{B}l}$ can be
considered to be a result of changes in $E_{\text{A}l}$ and $E_{0l}$. Hence,
we need to consider only the interaction Hamiltonian specified by (\ref%
{AppB_H}). The solution of the Schr\"{o}dinger equation for Hamiltonian (\ref%
{HsHint}) with (\ref{H0H1}) and (\ref{AppB_H}) (which can be validated by
substitution) is given here without derivation 
\begin{equation}
\left[ 
\begin{array}{c}
\tilde{A} \\ 
\tilde{B}%
\end{array}%
\right] _{l}=e^{-i(\omega _{0}+\omega _{_{l}}/2)t^{\circ }}\left[ 
\begin{array}{c}
\frac{b_{0}-b_{1}}{2b_{0}}\exp \left( i\frac{b_{0}t^{\circ }}{2}\right) +%
\frac{b_{0}+b_{1}}{2b_{0}}\exp \left( -i\frac{b_{0}t^{\circ }}{2}\right) \\ 
-2i\frac{\xi }{1+\xi ^{2}}\frac{\Delta \omega }{b_{0}}\sin \left( \frac{%
b_{0}t^{\circ }}{2}\right)%
\end{array}%
\right]  \label{AppB_AB2}
\end{equation}%
where 
\begin{equation}
b_{1}=+\Delta \omega \frac{1-\xi ^{2}}{1+\xi ^{2}}+\omega _{_{l}},\ \ \
b_{0}^{2}=\Delta \omega ^{2}+2\omega _{_{l}}\Delta \omega \frac{1-\xi ^{2}}{%
1+\xi ^{2}}+\omega _{_{l}}^{2},\ \ \ \omega _{_{l}}=\frac{E_{\text{A}l}}{%
\hbar }
\end{equation}%
while the other quantities $\omega _{0},$ $\Delta \omega $ and $\xi $ are
the same as defined in (\ref{A5eq1}), (\ref{A5eq2}) and (\ref{A5AB}). \ When 
$\omega _{_{l}}\ll \Delta \omega ,$ (\ref{AppB_AB2}) yields (\ref{ABt}) and
the interferences do not exercise much influence on the system. If $\omega
_{_{l}}\gg \Delta \omega $, these influences are strong since the asymptotic
limit of (\ref{AppB_AB2}) is given by 
\begin{equation}
\left[ 
\begin{array}{c}
\tilde{A} \\ 
\tilde{B}%
\end{array}%
\right] _{l}=e^{-i\omega _{0}t^{\circ }}\left[ 
\begin{array}{c}
\exp (-i\omega _{_{l}}t)+... \\ 
-\frac{\xi }{1+\xi ^{2}}\frac{\Delta \omega }{\omega _{_{l}}}(1-\exp
(-i\omega _{_{l}}t))%
\end{array}%
\right]
\end{equation}%
As in the previous subsection, tunnelling is accelerated by the factor of $%
2\omega _{_{l}}/\Delta \omega $, but the extent of tunnelling $\varsigma
=\left\vert \xi \right\vert /(1+\xi ^{2})$ is reduced by the factor of $%
\Delta \omega /\omega _{_{l}}$. At the limit of $\Delta \omega /\omega
_{l}\rightarrow 0,$ sections A and B become effectively isolated from each
other. Tracing out the degrees of freedom associated with the environment
suppresses the off-diagonal elements of the effective density matrix of the
system $\mathbf{\rho} _{s}$ but would not affect our conclusions limiting
the amplitude of $(\rho _{s})_{\text{BB}}.$

The extent of tunnelling can be enhanced by assuming that $\left\vert E_{%
\text{AB}l}\right\vert \neq 0,$ where$\ E_{\text{AB}l}=\left\langle \text{A}%
\right\vert \left\langle l\right\vert \mathbb{H}_{\text{int}}\left\vert 
\text{B}\right\rangle \left\vert l\right\rangle .$ This assumption, however,
does not seem physical, since it implies a rather strange possibility of
tunnelling from A to B through the environment (even if the magnitude of the
barrier $\hat{s}$ is prohibitively high to permit tunnelling).

\renewcommand{\emph} [1] {\textit{#1}}

\bibliographystyle{unsrtnat}
\bibliography{Law3}

\bigskip

\end{document}